\documentclass[a4paper,twocolumn, nofootinbib,aps,prl,eqsecnum,notitlepage,showkeys]{revtex4-2}
\usepackage{multirow}
\usepackage{graphicx}
\usepackage{bm}
\usepackage{array}
\usepackage{comment}
\usepackage{dcolumn}
\usepackage{hyperref}
\usepackage{gensymb}
\usepackage{amsmath}
\usepackage{dcolumn}    
\usepackage{ifpdf}
\usepackage{amssymb,lineno,amsfonts}
\usepackage{graphicx}   
\usepackage{bm}         
\usepackage{bbm}
\usepackage{mathrsfs}
\usepackage{upgreek}
\usepackage{mathtools}
\usepackage{epstopdf}
\usepackage{setspace}
\usepackage{hyperref}
\usepackage{natbib}
\usepackage{multirow}
\usepackage{esvect}
\usepackage{soul}
\usepackage{amsmath}
\usepackage[usenames,dvipsnames]{xcolor}
\definecolor{med-blue}{RGB}{0,78,255}
\hypersetup{colorlinks, linkcolor={blue},citecolor={Blue}, urlcolor={blue}}

\usepackage [english]{babel}
\usepackage [autostyle, english = american]{csquotes}
\MakeOuterQuote{"}
\begin{document}
	\title{
		Huge Symmetric Elastoresistance in the Kondo Lattice YbRh$_2$Si$_2$
	}
	\author{Soumendra Nath Panja}
	\email{soumendra.panja@uni-a.de}
	\address{Experimental Physics VI, Center for Electronic Correlations and Magnetism,\\
		University of Augsburg, 86159 Augsburg, Germany}
	\author{Jacques G Pontanel}
	\affiliation{Experimental Physics VI, Center for Electronic Correlations and Magnetism,\\
		University of Augsburg, 86159 Augsburg, Germany}
	\author{Julian Kaiser}
	\affiliation{Experimental Physics VI, Center for Electronic Correlations and Magnetism,\\
		University of Augsburg, 86159 Augsburg, Germany}
	\author{Anton Jesche}
	\affiliation{Experimental Physics VI, Center for Electronic Correlations and Magnetism,\\
		University of Augsburg, 86159 Augsburg, Germany}
	\author{Philipp Gegenwart}
	\email{philipp.gegenwart@uni-a.de}
	\affiliation{Experimental Physics VI, Center for Electronic Correlations and Magnetism,\\
		University of Augsburg, 86159 Augsburg, Germany}
	\date{\today}
	\begin{abstract}
		Heavy-fermion metals are prototype correlated electron systems for the study of Kondo entanglement and quantum criticality. We use the symmetry decomposed elastoresistance to uncover the fingerprints of strain-dependent Kondo scattering as function of temperature and magnetic field in the prototypical tetragonal Kondo lattice YbRh$_2$Si$_2$. By combining longitudinal and transverse resistance measurements under uniaxial strain applied along the tetragonal $[100]$ and $[110]$ directions, we obtain the elastoresistive responses in the in plane $A_{1g}$, $B_{1g}$, and $B_{2g}$ symmetry channels. While the responses in the symmetry-breaking channels are negligible, a huge enhancement of the isotropic symmetric $A_{1g}$ elastoresistance is found approaching -50 at low temperatures. Scaling analysis and comparison with linear thermal expansion measurements reveal that the symmetric elastoresistance probes the contribution of Kondo scattering to the strain dependence of magnetic entropy and signals strain- and field controlled quantum criticality upon cooling to 2~K.
		
	\end{abstract} 
	\pacs{Pacs}
	\maketitle
	$\textit{Introduction}$---Uniaxial lattice strain has emerged as a uniquely powerful, symmetry-selective tuning parameter for correlated electron systems, providing direct control over electronic energy scales without introducing chemical disorder, as long as the response remains in the elastic limit of the studied material~\cite{HicksARCMP,Jo2024, Wang2023,Lieberich2025}. 
The elastoresistance quantifies the change of resistance with strain.
When analyzed in the irreducible representations of the lattice point group, it offers a microscopic probe of how distinct symmetry components of the lattice couple to electronic degrees of freedom. Understanding how lattice symmetry governs the interplay between electronic/magnetic energy scales is a central theme in correlated electron physics. This approach uncovered divergent nematic susceptibilities in iron-based superconductors \cite{ScienceJHChu,Kuo,Hosoi,Ishida, PRLWiecki, Wiecki2021,Worasaran} and revealed symmetry-selective couplings in a range of correlated electron materials~\cite{Rosenberg2019,Ye2023,Ye2024,Rosenberg2024}.
 
Below, we report the first study of the symmetry-resolved elastoresistance in a heavy-fermion metal. This material class is prototypical for studying correlated electron behavior, emergent orders, quantum criticality and unconventional superconductivity~\cite{Gegenwart2008}. Heavy-fermion metals \cite{Stewart1984} are traditionally discussed within the Doniach phase diagram \cite{Doniach1977}, where the ground state is determined by the competition between Kondo screening of localized 4$f$ moments and the RKKY interaction that promotes magnetic order. The antiferromagnetic (AFM) exchange between local moments and conduction electrons controlling both interactions leads to exceptionally large Gr\"uneisen parameters \cite{Gegenwart2016}, highlighting the strong sensitivity of these materials to pressure.

Fig.~\ref{figure1} sketches the generic temperature and field dependence of the electrical resistance of Kondo lattices by the black lines in the two panels.
As shown in (a) the electrical resistance increases upon cooling due to the Kondo effect, passes a maximum, followed by a decrease to low temperatures, due to the formation of the correlated quasiparticle bands. This generic temperature dependence scales with the Kondo temperature $T_{\rm K}$~\cite{Lavagna1982,Coleman1987,Cox1988}. In the case of crystal electric field (CEF) splitting this refers to the Kondo interaction energy with the full multiplett, which exceeds that for the lowest lying CEF level~\cite{Cox1988}. Due to the polarization of Kondo singlets in field, which leads to a reduction of the resistance, Kondo lattices often show a negative magnetoresistance below $T_{\rm K}$, which scales like $H/(T + T_{\rm K})$ and has a form as indicated by the black line in panel (b)~\cite{Schlottmann1983,Batlogg1987}. Assuming that compressive and tensile strain $\varepsilon$ reduces and enhances the Kondo temperature (as may be expected for the Yb case), this results in the green and blue curves, respectively direct elastoresistance, affecting the height of the resistance maxima $\rho_{\rm max}$, has been neglected). This allows to sketch the possible elastoresistance $(d\rho/d\varepsilon)/\rho$ behavior, see the red lines. Upon cooling, the elastoresistance might change its sign near $T_{\rm max}$ of the ambient $\rho(T)$ curve and upon further cooling decreases to enhanced negative values (panel a), while the application of magnetic fields at low temperatures may leads to a pronounced positive elastoresistance (panel b). To explore experimentally the elastoresistance fingerprints of Kondo lattices, we focus on YbRh$_2$Si$_2$ (YRS).

	
	\begin{figure}
		\includegraphics[scale=0.35]{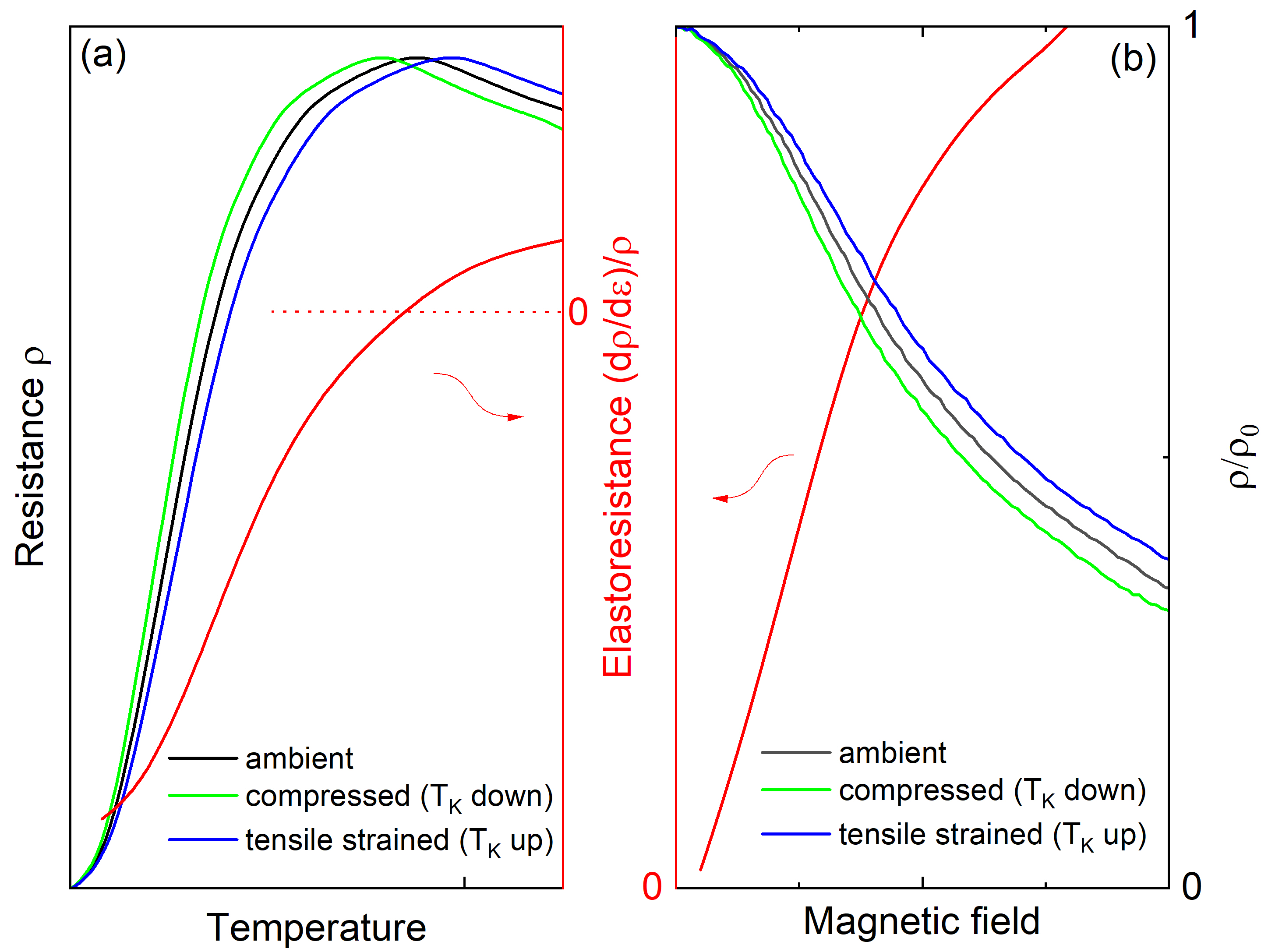}
		\caption{Schematic temperature (a) and field dependence (b) of the electrical resistance $\rho$ of the Kondo lattice~\cite{Cox1988,Schlottmann1983,Batlogg1987}, indicated by black lines. The blue and green lines indicate the effect of an increase and decrease of $T_{\rm K}$, respectively. The elastoresistance $(d\rho/d\varepsilon)/\rho$ (in red, with right y-axis in (a) and left y-axis in (b)) is estimated under the assumption of a decrease/increase of $T_{\rm K}$ under compression/tensile strain $\varepsilon$. Note the sign change in (a) and positive elastoresistance under field in (b).}
		\label{figure1}
	\end{figure}

	YRS crystallizes in the well-known tetragonal ThCr$_2$Si$_2$ structure. It is a prototypical heavy fermion system 
	 proximate to an AFM quantum critical point (QCP), related to very weak AFM order at $T_{\rm N} \approx 70$~mK~\cite{Gegenwart2002}. Hydrostatic pressure enhances $T_{\rm N}$ and its critical magnetic field~\cite{PRLTokiwa}. A zero-field QCP cannot be further approached, since this would require a volume expansion beyond what is possible by partial isoelectronic substitution of Si by the larger Ge in YbRh$_2$(Si$_{1-x}$Ge$_x$)$_2$~\cite{Custers2003}. So far approaching the QCP requires the application of magnetic field, which however polarizes the ferromagnetic component of critical fluctuations~\cite{Gegenwart2005}. Previously, we reported the impact of $a$-axis tensile and compressive strain between $-1.7\cdot 10^{-3}$ and $+1.3\cdot 10^{-3}$ on the resistivity maximum of YRS by electrical resistance measurements in a piezoelectric force cell (type FC100 from Razorbill) between 100 and 160~K, which revealed an increase of $T_{\rm max}$ with tensile strain~\cite{Panja2024}. However, that study did neither investigate the temperature dependence of the elastoresistance down to low temperatures nor its anisotropy and decomposition in the contributions arising from symmetry conserving and breaking strain channels which are essential for characterizing the impact of strain on Kondo lattices.
	
	
In this Letter, we address this question through comprehensive symmetry-resolved elastoresistance measurements on tetragonal YRS. By combining longitudinal and transverse measurements under uniaxial strain applied along the [100] and [110] directions, we extract the complete elastoresistive response in the in plane $A_{1g}$, $B_{1g}$, and $B_{2g}$ symmetry channels for the first time in a Kondo lattice metal. We find that the response is entirely dominated by the isotropic $A_{1g}$ channel, demonstrating that in-plane strain tunes the Kondo hybridization, with no detectable contribution from either symmetry-breaking channels. The observed temperature and field dependences are in accordance with the sketches of Fig.~\ref{figure1} discussed above and highlight the enormous strain dependence of the electrical resistance of Kondo lattices. Comparative measurements of the linear thermal expansion coefficient indicate that the elastoresistive scattering response tracks the strain dependence of the magnetic entropy and as such is a novel sensitive probe to characterize heavy-fermion behavior.


\begin{figure}
		\includegraphics[scale=0.30]{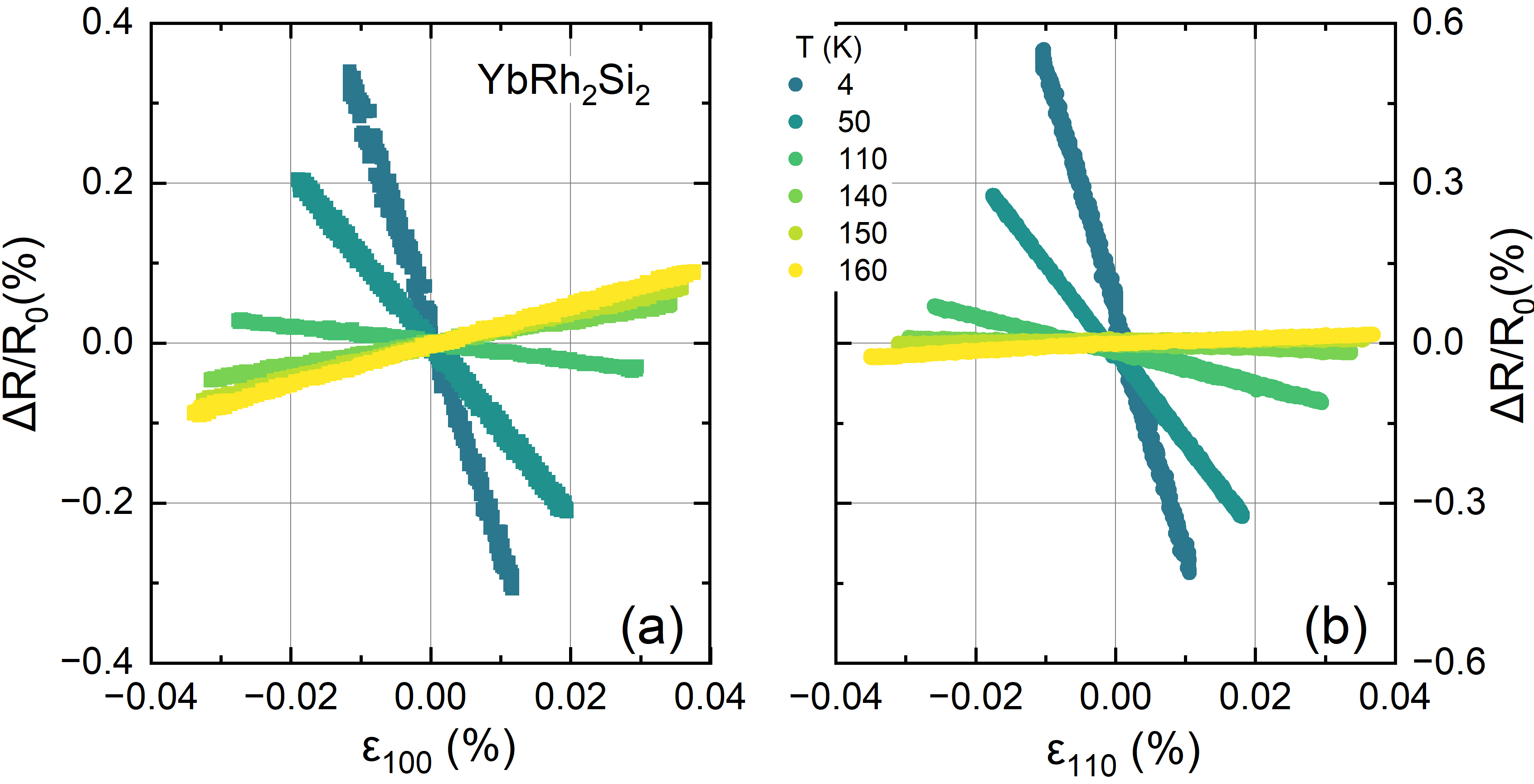}
		\caption{Longitudinal strain-induced change in electrical resistance $\Delta R/R_0$ for YRS at various temperatures for strain and current parallel to the $[100]$ (a) and $[110]$ (b) directions. $R_0$ denotes the resistance without applied strain.}
		\label{figure2}
	\end{figure}

		$\textit{Methods}$--- 		
				YRS single crystals were prepared and treated as described in ~\cite{Panja2024}. Elastoresistance measurements were performed using a commercial strain cell (Razorbill Instruments CS-100) \cite{Hicks2014Rsi}. In this setup, the sample is mounted and suspended between two movable plates, with their separation precisely controlled by the applied voltage on piezoelectric stacks. Plate-like single crystals were oriented and cut with their edges in the tetragonal in-plane either along [100] or [110], with typical dimensions of approximately $2.0 \times 0.2~ \mathrm{to}~ 0.6 \times 0.05~\mathrm{mm^3}$ where the shortest direction is along [001]. The samples were bonded to the strain cell using Stycast 2850FT, resulting in an effective strained length of approximately 0.8 to 1~mm. The applied strain was monitored using a built-in capacitive displacement sensor. Four-probe electrical resistance measurements as function of temperature and strain were carried out in a PPMS (Quantum Design), with the strain cell mounted on a modified P450 probe (Quantum Design). Thermal anchoring of the cell was achieved using silver foil and wires. 
	Eight-contact configurations (as shown in Fig.~S1 \cite{supple}) were employed for current flow either longitudinal or transverse to the direction of applied strain. This allows to measure simultaneously the longitudinal and transverse electrical resistance on each crystal at low temperature in magnetic fields applied parallel to strain direction.
	To determine the elastoresistance, isothermal sweeps of the piezo-voltage were utilized, recording the strain sensor and the electrical resistance, simultaneously.

	 $\textit{Results and discussion}$--- Fig.~\ref{figure2} shows the normalized longitudinal resistance as function of strain as $\Delta R$/$R_0$ with $\Delta R=R-R_0$ where $R_0$ denotes the resistance without applied strain for small uniaxial strain $\varepsilon$ applied along the in-plane crystalline directions [100] and [110] in panels (a) and (b) respectively. Importantly all curves are linear in $\varepsilon$ without noticeble hysteresis, indicating fully elastic behavior.
	 

	 	The strain response can be expressed through the strain derivative of the relative resistance change,
	 	\( d(\Delta R/R_{0})/d\varepsilon \), which has been calculated fitting the linear strain response of the curves plotted in Fig.~\ref{figure2} for all measured temperatures. Generically, this property is the sum of a geometric and the intrinsic resistivity (\(\rho\)) contributions.
	 	Using \( R = \rho L/A \), the geometric part reduces to \( 1 + 2\nu \) (cf. eq. S1 in \cite{supple}) with Poisson's ratio $\nu$ typically lying in the range \( \nu \approx 0.25\text{--}0.35 \)
	 	(an upper theoretical limit for any stable isotropic solid is \( \nu = 0.5 \)). This yields a geometric contribution of approximately \( 1.5\text{--}1.8 \) \cite{Callister2018, Sun2010Strain} (with a maximum possible value of 2).
	 	Any measured strain derivative that substantially differs from this baseline therefore reflects the intrinsic strain dependence of the resistivity.
	 	
	 	\begin{figure}
		\includegraphics[scale=0.26]{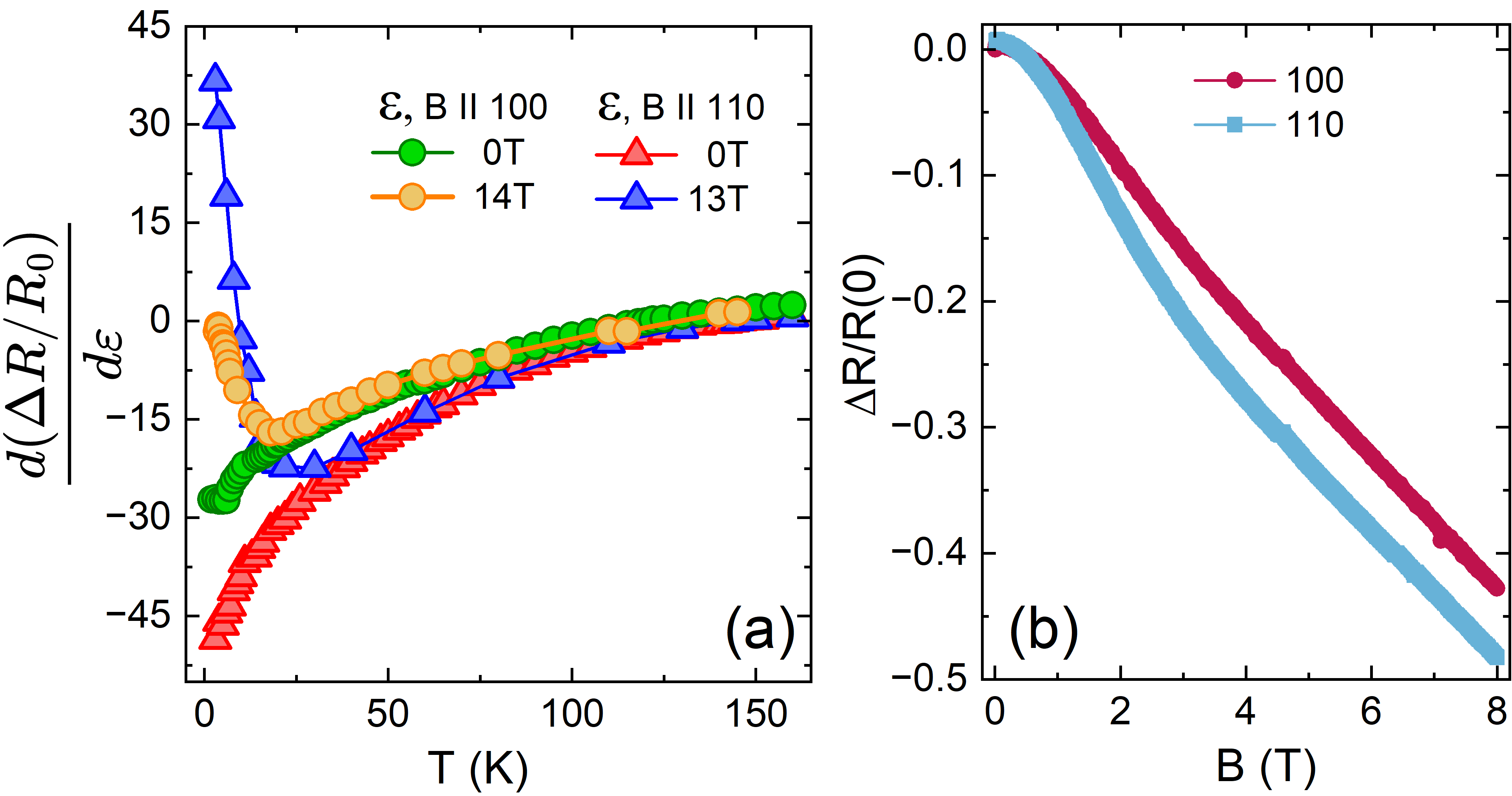}
		\caption{a): Temperature dependence of the relative resistance change under strain $d(\Delta R/R_0)/d\varepsilon$ along the [100] and [110] directions,  at zero field and under applied magnetic fields.  (b): Magnetoresistance $\Delta R(B)/R(0)$ for current along the [100] and [110] directions at 2 K.}
		\label{figure3}
	\end{figure}

Figure~\ref{figure3} summarizes the slope of resistance changes for uniaxial strain applied along the $[100]$ and $[110]$ directions. At high temperatures around 150 K, the values $\Delta R/R_{0}$ are almost isotropic and field independent, suggesting a dominating geometrical contribution due to the Poisson effect. The sign change for the data along the $[100]$ direction near 120 K to negative values is consistent with our previous data plotted in Fig. 4(b) of ~\cite{Panja2024} and in qualitative agreement with the expectation for Yb-based Kondo lattices, sketched in Fig.~\ref{figure1}. The same holds for the increasingly larger negative values upon further cooling. This negative contribution is larger along $[110]$ but our symmetry-resolved analysis presented below indicates an isotropic symmetric strain coupling. Interestingly, when a magnetic field is applied, $d(\Delta R/R_0)/d\varepsilon$ deviates from its zero-field trend below about 40~K and displays a increase upon cooling~\cite{supple}. A positive contribution to elastoresistance in magnetic field at temperatures below $T_{\rm K}$ is consistent with the strain effect on the Kondo magnetoresistance, sketched in Fig.~\ref{figure1}(b). Below about 10~K, where zero-field ambient pressure electrical resistance of YRS enters the $T$-linear regime characteristic for the onset of quantum criticality, the positive elastoresistance contribution in field further grows.
Overall the data demonstrate the power of elastoresistance to probe the strain dependence of Kondo-lattice interactions. 

\begin{figure}
		\includegraphics[scale=0.52]{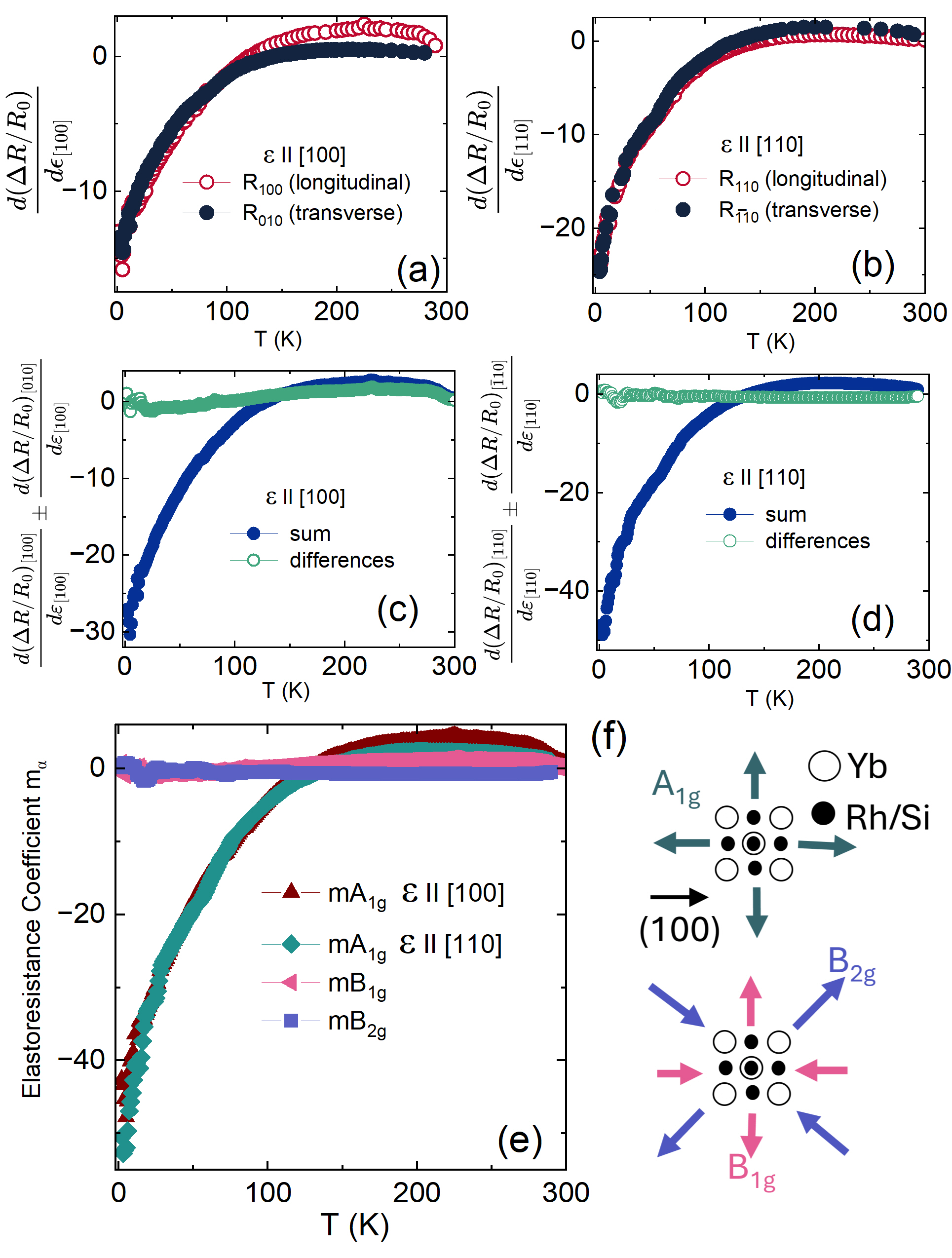}
		\caption{Symmetry decomposition of elastoresistance. (a) and (b) depict the strain derivatives of the longitudinal and transverse resistance for uniaxial strain applied along the [100] and [110] directions, respectively. (c) and (d) depict the sum and difference of longitudinal and transverse strain derivatives along [100] and [110], respectively. (e) Temperature dependence of the elastoresistance coefficients $m_{A1g}$, $m_{B1g}$, and $m_{B2g}$, calculated from the data in (c) and (d) using 
eqs. (S2)-(S5)~\cite{supple}. (f) Schematic illustration of $A_{1g}$, $B_{1g}$, and $B_{2g}$ strain symmetry channels in YRS.}
		\label{figure4}
	\end{figure}

	 Symmetry-resolved strain measurement were performed to further characterize the elastoresistive responses. In a tetragonal crystal such as YRS, the in-plane elastic strain can be decomposed into distinct symmetry channels of the $D_{4h}$ point group as schematically depicted in Fig.~\ref{figure4}(f). 
	An \textit{isotropic} ($A_{1g}$) strain changes the in-plane area without breaking the fourfold ($C_4$) rotational symmetry.
	By contrast,  $B_{1g}$ and $B_{2g}$ strains both break the $C_4$ symmetry and are therefore sensitive to nematicity. The $B_{1g}$ channel corresponds to an \textit{orthorhombic distortion}, characterized by stretching one axis while compressing the perpendicular one, $\varepsilon_{B1g} \sim (\varepsilon_{xx} - \varepsilon_{yy})$, whereas the $B_{2g}$ channel represents a \textit{shear distortion}, $\varepsilon_{B2g} \sim \varepsilon_{xy}$, which skews the lattice along the $[110]$ and [1$\bar{1}$0] diagonals. When uniaxial strain is applied along the $[100]$ (or $[110]$) direction, the sample experiences $\varepsilon_{A1g+ B1g}$ (or $\varepsilon_{A1g + B2g}$) strain. For strain applied along the $[100]$ direction, the longitudinal deformation $\varepsilon$ is accompanied by a transverse contraction $-\nu\,\varepsilon$, where $\nu$ is the Poisson ratio. This produces symmetry strains $\varepsilon_{A1g} = \tfrac{1}{2}(1-\nu)\,\varepsilon$ and $\varepsilon_{B1g} = \tfrac{1}{2}(1+\nu)\,\varepsilon$, with no shear component 
	 in the $B_{2g}$ channel. Conversely, for strain applied along the $[110]$ direction, 
	 the transverse response $-\nu\,\varepsilon$ generates 
	 $\varepsilon_{A1g} = \tfrac{1}{2}(1-\nu)\,\varepsilon$ and a finite shear strain 
	 $\varepsilon_{B2g} = \tfrac{1}{2}(1+\nu)\,\varepsilon$, while the $B_{1g}$ 
	 component vanishes. The intrinsic elastoresistivity response is defined as $m_{\Gamma}(T,B) = \frac{1}{\rho}\frac{\partial \rho}{\partial \varepsilon_{\Gamma} }$, where $\Gamma = A_{1g},\, B_{1g},\, B_{2g}$. The sum of the longitudinal and transverse resistance changes isolates the $A_{1g}$ channel, while their difference isolates the symmetry-breaking $B_{1g}$ (for applied strain along $[100]$) or $B_{2g}$ (for applied strain along $[110]$) channel. The in-plane anisotropic Poisson's ratios $\nu_{[100]} = -\varepsilon_{[010]}/\varepsilon_{[100]}$ and $\nu_{[110]} = -\varepsilon_{[\bar{1}10]}/\varepsilon_{[110]}$ relate longitudinal and transverse strain components. The explicit relations used to extract the symmetry-resolved elastoresistivity coefficients are given in Eqs. (S2)-(S5)~\cite{supple}.
	Poisson's ratios were determined from elastic constants computed via density-functional theory (GGA); the resulting temperature-independent values are $\nu_{[100]} \approx 0.36$ and $\nu_{[110]} \approx 0.07$ \cite{supple}.
	
  To resolve the strain response into its symmetry channels, we measured the longitudinal and transverse elastoresistance using an eight-contact configuration, with four contacts allocated to each direction. Measurements were performed separately for the $[100]$ and $[110]$ orientations.
 Fig.\ref{figure4} (a) and (b) show the resulting slopes of relative change in resistance $d(\Delta R/R_{0})/d\varepsilon$ for current aligned parallel and perpendicular to the applied uniaxial strain along [100] and [110]. For both strain directions, the longitudinal and transverse responses follow one another closely over the entire temperature range, indicating only 
	weak anisotropic response. Fig.~\ref{figure4}(c) and (d) display the corresponding symmetry combinations, i.e. the sums and differences for the $[100]$ and $[110]$ orientations, respectively. These combinations of measurements isolate the couplings of the electronic system to strain in the three irreducible representations of the tetragonal point group from the geometric contribution.
		The extracted symmetry-resolved coefficients $m_{A1g}$, $m_{B1g}$, and $m_{B2g}$ are shown in Fig. \ref{figure4}(e). Importantly, both $m_{B1g}$ and $m_{B2g}$ remain negligible within experimental resolution upon cooling to low temperatures, demonstrating the absence of response in symmetry breaking channels.
		
		On the other hand, a large negative contribution is observed in the $A_{1g}$ channel with growing magnitude upon decreasing temperature. Remarkably the $m_{A1g}$ data are similar when the coefficient is extracted from the $[100]$ or $[110]$ elastoresistance, indicating that the results are robust against systematic errors. At 2~K an enhanced value near $-50$ is found, whose absolute magnitude is similar to that found in hole-doped iron arsenides, which are close to the orbital-selective Mott insulator transition~\cite{Wiecki2021}. However, there are striking differences in YRS: first, the sign of $m_{A1g}$ is negative, which is related to the pressure dependence of $T_{\rm K}$ for Yb-based heavy-fermion metals as discussed above. Second, the strong temperature dependence of $m_{A1g}$ down to the 4~K range suggests that the symmetric elastoresistance captures non-Fermi liquid behavior and the onset of quantum criticality in YRS, while in case of the most correlated iron pnictide CsFe$_2$As$_2$ Fermi liquid behavior is found below 10~K~\cite{Wiecki2021}.		
				
		
Since the resistivity at low temperatures is dominated by short-range magnetic scattering we may apply the Fisher-Langer scaling relation $d\rho/dT\sim C$ between the magnetic specific heat and the temperature derivative of the resistance~\cite{Fisher1968} and take the strain derivative on both sides in order to further analyse the elastoresistance behavior. As detailed~\cite{supple}, this provides a relation between elastoresistance and the linear thermal expansion $\alpha_i=1/L_i(d\Delta L_i/dT)$ (where $L_i(T)$ is the sample length along the measured direction), which probes the $\epsilon_i$ strain derivative of entropy~\cite{Gegenwart2016}. More precisely, the scaling relation

		\begin{equation}
		\frac{\alpha_i}{T} \propto -\frac{1}{T}\int \left[ 
		\frac{1}{T}\frac{\partial}{\partial T}\!\left( 
		\frac{\partial \rho}{\partial p_i} 
		\right)\right] dT,
		\tag{1}
	\end{equation}
	
can be derived~\cite{Wiecki2021}. For the case of symmetric in-plane stress we use $(\partial \rho/\partial p_{A1g})=-c_{A1g}^{-1}(\partial \rho/\partial \varepsilon_{A1g})= -m_{A1g} R(T)/R(300 \rm K)\cdot \rho(300 \rm K)$, with the elastic constant for symmetric in-plane stress $c_{A1g}$. The negative sign follows from the convention that pressure is positive under compression, whereas strain is positive under tension. Inserting to Eq.~(1) yields
	\begin{equation}
		\frac{\alpha_i}{T} \propto \frac{1}{T}\int 
		\frac{1}{T}\frac{\partial}{\partial T}\!\left( 
		\frac{m_{A1g}R(T)}{R(300 \rm K)}
		\right) dT,
		\tag{2}
		\label{2}
	\end{equation}
which relates the symmetric-strain elastoresistance $m_{A1g}$ to the magnetic contribution of the expansion coefficient~\cite{Wiecki2021} and thus allows to check, how the elastoresistance reflects the strain dependence of  magnetic entropy. For this purpose, we utilized a high-resolution capacitive dilatometer with a modified sample stage for small and very thin single crystals~\cite{Panja2025} to measure the linear thermal expansion of YRS	along both the $[100]$ and $[110]$ direction at temperatures between 2 and 200 K.

Earlier in-plane thermal expansion measurements between 50 mK and 6 K showed a negative linear in-plane thermal expansion coefficient above $T_{\rm N}$~\cite{Kuechler2004}. The negative sign of the (divergent) Gr\"uneisen parameter indicates the decrease of $T_{\rm K}$ under pressure~\cite{Kuechler2003}. In the overlap regime from 2 to 6 K, our data agree well with the previous low-$T$ experiment.

	\begin{figure}
		\includegraphics[width=\columnwidth]{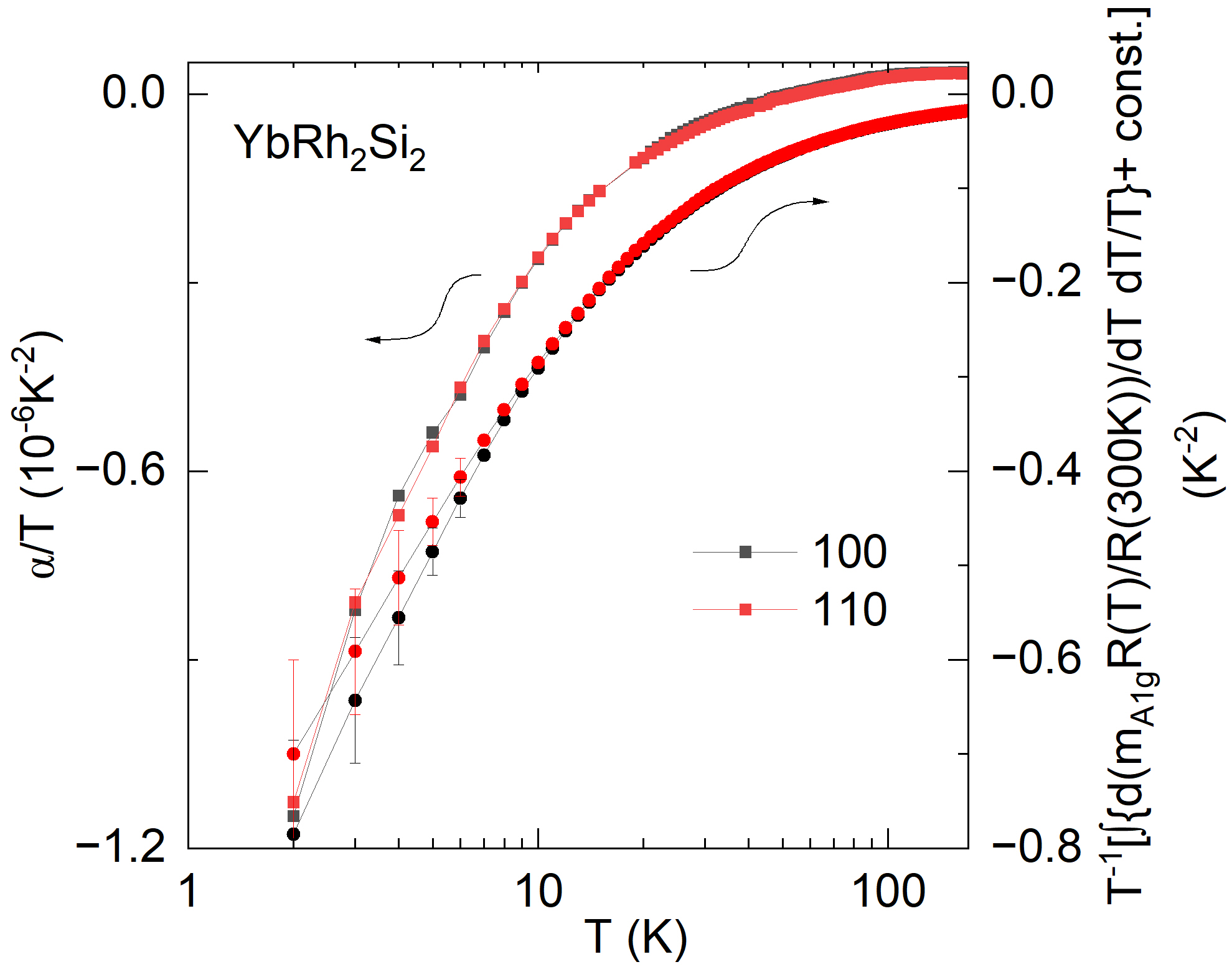}
		\caption{In-plane thermal expansion coefficient as $\alpha(T)/T$ vs $T$ (on log scale) for YRS measured along the [100] and [110] directions, indicated by red and black squares, respectively (left y-axis). The red and black circles display the respective values of $T^{-1}\int[(dT/T)(d(m_{A1g}R/R_{300\rm K})/dT)+const.]$ (right y-axis) with the integration constant as determined in~\cite{supple}.}
		\label{figure5}
	\end{figure}

In order to test the scaling relation Eq. (2), we now compare the temperature dependences of the thermal expansion coefficient as $\alpha/T$ with the respective integrated temperature derivative of elastoresistance in	Fig.~\ref{figure5} for the [100] and [110] directions. As detailed in supplemental material and Fig. S5, an integration constant related to the unknown elastoresistance below 2~K has been added~\cite{supple}. The similarity between $\alpha/T$ and the respective elastoresistance derived quantity suggests that elastoresistance in Kondo lattices probes the strain dependence of magnetic entropy, which for YRS is enhanced due to the proximity to quantum criticality~\cite{Kuechler2003}.

	
	
Entropy accumulation near a QCP that is realized by variation of a control parameter $r$ results in a generic {\it divergence} of the respective Gr\"uneisen parameter $\Gamma_r={1\over T}\cdot {dT\over dr}|_S$ ~\cite{Zhu2003}. Adiabatic magnetocaloric ($r=B$) and elastocaloric ($r=\epsilon$) effects, which can be detected by high-resolution ac techniques~\cite{Tokiwa,Li2022}, are thus ideal probes for the study and characterization of QCPs. The Fisher-Langer derived relation between thermal expansion and elastoresistance, shown in Fig.~\ref{figure5} indicates that elastoresistance also captures the strain dependent magnetic entropy, though indirectly via the scattering of conduction electrons. This opens a perspective for future comparative studies of the elastocaloric effect and elastoresistance. Electrical transport can distinguish quantum criticality scenarios where only a small fraction of the Fermi surface is affected or singular scattering at the entire Fermi surface in case of local, q-independent, criticality~\cite{Hlubina1995,Coleman2001}. Furthermore, in case of Mott criticality, electrical transport directly probes the evolution of the charge gap~\cite{Pustogow2023}. It is therefore obvious that elastoresistance can be a powerfull novel tool for the study of quantum criticality.


$\textit{Conclusions}$---	We have performed symmetry-resolved elastoresistance measurements on the archetypal heavy-fermion metal YbRh$_2$Si$_2$ to determine how in-plane strain couples to its low-energy electronic and magnetic properties. A huge negative elastoresistance with a positive field-induced contribution is found at low temperatures. This behavior, illustrated in Fig.~\ref{figure1}, shall be generic for Yb-based Kondo lattices. Longitudinal and transverse electrical resistance measurements under uniaxial strain along $[100]$ and $[110]$ confirmed the above qualitative expectation and revealed the quantitative elastoresistive response which is overwhelmingly dominated by the isotropic $A_{1g}$ channel, which upon cooling reaches huge values of order $-50$ already at 2~K. The absence of detectable $B_{1g}$ and $B_{2g}$ responses shows that in-plane strain exclusively tunes the overall Kondo hybridization without inducing symmetry-breaking electronic and structural distortions. The close correspondence between the $A_{1g}$ elastoresistance and the in-plane thermal expansion establishes a direct link to the strain dependence of the magnetic entropy, highlighting elastoresistance as a novel tool to probe the coupling of conduction electrons to the strain dependent magnetic correlations in heavy-fermion metals. Previously, we demonstrated that tensile strain can tune $T_{\rm K}$ of YRS to the value required for approaching the extrapolated zero-field quantum critical point. Symmetry-resolved elastoresistance measurements at large tensile strain and down to the milli-Kelvin range are thus of huge interest to uncover the nature of its quantum critical state, avoiding additional complexity associated with magnetic field polarization of ferromagnetic fluctuations near the critical field or disorder in chemically substituted YRS~\cite{Gegenwart2005,Schubert}.

	$\textit{Acknowledgments}$---We acknowledge fruitful discussions with A. E. Böhmer and G. Zwicknagl. S.N.P. was supported by the Alexander von Humboldt Foundation.
	\bibliography{Bibliography}

\begin{thebibliography}{46}%
\makeatletter
\providecommand \@ifxundefined [1]{%
 \@ifx{#1\undefined}
}%
\providecommand \@ifnum [1]{%
 \ifnum #1\expandafter \@firstoftwo
 \else \expandafter \@secondoftwo
 \fi
}%
\providecommand \@ifx [1]{%
 \ifx #1\expandafter \@firstoftwo
 \else \expandafter \@secondoftwo
 \fi
}%
\providecommand \natexlab [1]{#1}%
\providecommand \enquote  [1]{``#1''}%
\providecommand \bibnamefont  [1]{#1}%
\providecommand \bibfnamefont [1]{#1}%
\providecommand \citenamefont [1]{#1}%
\providecommand \href@noop [0]{\@secondoftwo}%
\providecommand \href [0]{\begingroup \@sanitize@url \@href}%
\providecommand \@href[1]{\@@startlink{#1}\@@href}%
\providecommand \@@href[1]{\endgroup#1\@@endlink}%
\providecommand \@sanitize@url [0]{\catcode `\\12\catcode `\$12\catcode
  `\&12\catcode `\#12\catcode `\^12\catcode `\_12\catcode `\%12\relax}%
\providecommand \@@startlink[1]{}%
\providecommand \@@endlink[0]{}%
\providecommand \url  [0]{\begingroup\@sanitize@url \@url }%
\providecommand \@url [1]{\endgroup\@href {#1}{\urlprefix }}%
\providecommand \urlprefix  [0]{URL }%
\providecommand \Eprint [0]{\href }%
\providecommand \doibase [0]{https://doi.org/}%
\providecommand \selectlanguage [0]{\@gobble}%
\providecommand \bibinfo  [0]{\@secondoftwo}%
\providecommand \bibfield  [0]{\@secondoftwo}%
\providecommand \translation [1]{[#1]}%
\providecommand \BibitemOpen [0]{}%
\providecommand \bibitemStop [0]{}%
\providecommand \bibitemNoStop [0]{.\EOS\space}%
\providecommand \EOS [0]{\spacefactor3000\relax}%
\providecommand \BibitemShut  [1]{\csname bibitem#1\endcsname}%
\let\auto@bib@innerbib\@empty
\bibitem [{\citenamefont {Hicks}\ \emph {et~al.}(2025)\citenamefont {Hicks},
  \citenamefont {Jerzembeck}, \citenamefont {Noad}, \citenamefont {Barber},\
  and\ \citenamefont {Mackenzie}}]{HicksARCMP}%
  \BibitemOpen
  \bibfield  {author} {\bibinfo {author} {\bibfnamefont {C.~W.}\ \bibnamefont
  {Hicks}}, \bibinfo {author} {\bibfnamefont {F.}~\bibnamefont {Jerzembeck}},
  \bibinfo {author} {\bibfnamefont {H.~M.}\ \bibnamefont {Noad}}, \bibinfo
  {author} {\bibfnamefont {M.~E.}\ \bibnamefont {Barber}},\ and\ \bibinfo
  {author} {\bibfnamefont {A.~P.}\ \bibnamefont {Mackenzie}},\ }\bibfield
  {title} {\bibinfo {title} {Probing quantum materials with uniaxial stress},\
  }\href
  {https://doi.org/https://doi.org/10.1146/annurev-conmatphys-040521-041041}
  {\bibfield  {journal} {\bibinfo  {journal} {Annu. Rev. Condens. Matter
  Phys.}\ }\textbf {\bibinfo {volume} {16}},\ \bibinfo {pages} {417} (\bibinfo
  {year} {2025})}\BibitemShut {NoStop}%
\bibitem [{\citenamefont {Hyun~Jo}\ \emph {et~al.}(2024)\citenamefont
  {Hyun~Jo}, \citenamefont {Gati},\ and\ \citenamefont {Pfau}}]{Jo2024}%
  \BibitemOpen
  \bibfield  {author} {\bibinfo {author} {\bibfnamefont {N.}~\bibnamefont
  {Hyun~Jo}}, \bibinfo {author} {\bibfnamefont {E.}~\bibnamefont {Gati}},\ and\
  \bibinfo {author} {\bibfnamefont {H.}~\bibnamefont {Pfau}},\ }\bibfield
  {title} {\bibinfo {title} {{Uniaxial stress effect on the electronic
  structure of quantum materials}},\ }\href
  {https://doi.org/10.3389/femat.2024.1392760} {\bibfield  {journal} {\bibinfo
  {journal} {Front. Electron. Mater.}\ }\textbf {\bibinfo {volume} {4}},\
  \bibinfo {pages} {1392760} (\bibinfo {year} {2024})}\BibitemShut {NoStop}%
\bibitem [{\citenamefont {Wang}\ \emph {et~al.}(2023)\citenamefont {Wang},
  \citenamefont {Spitaler}, \citenamefont {Su}, \citenamefont {Zoch},
  \citenamefont {Krellner}, \citenamefont {Puphal}, \citenamefont {Brown},\
  and\ \citenamefont {Pustogow}}]{Wang2023}%
  \BibitemOpen
  \bibfield  {author} {\bibinfo {author} {\bibfnamefont {J.}~\bibnamefont
  {Wang}}, \bibinfo {author} {\bibfnamefont {M.}~\bibnamefont {Spitaler}},
  \bibinfo {author} {\bibfnamefont {Y.-S.}\ \bibnamefont {Su}}, \bibinfo
  {author} {\bibfnamefont {K.}~\bibnamefont {Zoch}}, \bibinfo {author}
  {\bibfnamefont {C.}~\bibnamefont {Krellner}}, \bibinfo {author}
  {\bibfnamefont {P.}~\bibnamefont {Puphal}}, \bibinfo {author} {\bibfnamefont
  {S.}~\bibnamefont {Brown}},\ and\ \bibinfo {author} {\bibfnamefont
  {A.}~\bibnamefont {Pustogow}},\ }\bibfield  {title} {\bibinfo {title}
  {{Controlled Frustration Release on the Kagome Lattice by Uniaxial-Strain
  Tuning}},\ }\href {https://doi.org/10.1103/PhysRevLett.131.256501} {\bibfield
   {journal} {\bibinfo  {journal} {Phys. Rev. Lett.}\ }\textbf {\bibinfo
  {volume} {131}},\ \bibinfo {pages} {256501} (\bibinfo {year}
  {2023})}\BibitemShut {NoStop}%
\bibitem [{\citenamefont {Lieberich}\ \emph {et~al.}(2025)\citenamefont
  {Lieberich}, \citenamefont {Saito}, \citenamefont {Agarmani}, \citenamefont
  {Sasaki}, \citenamefont {Yoneyama}, \citenamefont {Winter}, \citenamefont
  {Lang},\ and\ \citenamefont {Gati}}]{Lieberich2025}%
  \BibitemOpen
  \bibfield  {author} {\bibinfo {author} {\bibfnamefont {F.}~\bibnamefont
  {Lieberich}}, \bibinfo {author} {\bibfnamefont {Y.}~\bibnamefont {Saito}},
  \bibinfo {author} {\bibfnamefont {Y.}~\bibnamefont {Agarmani}}, \bibinfo
  {author} {\bibfnamefont {T.}~\bibnamefont {Sasaki}}, \bibinfo {author}
  {\bibfnamefont {N.}~\bibnamefont {Yoneyama}}, \bibinfo {author}
  {\bibfnamefont {S.~M.}\ \bibnamefont {Winter}}, \bibinfo {author}
  {\bibfnamefont {M.}~\bibnamefont {Lang}},\ and\ \bibinfo {author}
  {\bibfnamefont {E.}~\bibnamefont {Gati}},\ }\bibfield  {title} {\bibinfo
  {title} {{Probing and tuning geometric frustration in an organic quantum
  magnet via elastocaloric measurements under strain}},\ }\href
  {https://doi.org/10.1126/sciadv.adz0699} {\bibfield  {journal} {\bibinfo
  {journal} {Sci. Adv.}\ }\textbf {\bibinfo {volume} {11}},\ \bibinfo {pages}
  {eadz0699} (\bibinfo {year} {2025})}\BibitemShut {NoStop}%
\bibitem [{\citenamefont {Chu}\ \emph {et~al.}(2012)\citenamefont {Chu},
  \citenamefont {Kuo}, \citenamefont {Analytis},\ and\ \citenamefont
  {Fisher}}]{ScienceJHChu}%
  \BibitemOpen
  \bibfield  {author} {\bibinfo {author} {\bibfnamefont {J.-H.}\ \bibnamefont
  {Chu}}, \bibinfo {author} {\bibfnamefont {H.-H.}\ \bibnamefont {Kuo}},
  \bibinfo {author} {\bibfnamefont {J.~G.}\ \bibnamefont {Analytis}},\ and\
  \bibinfo {author} {\bibfnamefont {I.~R.}\ \bibnamefont {Fisher}},\ }\bibfield
   {title} {\bibinfo {title} {{Divergent Nematic Susceptibility in an Iron
  Arsenide Superconductor}},\ }\href {https://doi.org/10.1126/science.1221713}
  {\bibfield  {journal} {\bibinfo  {journal} {Science}\ }\textbf {\bibinfo
  {volume} {337}},\ \bibinfo {pages} {710} (\bibinfo {year}
  {2012})}\BibitemShut {NoStop}%
\bibitem [{\citenamefont {Kuo}\ \emph {et~al.}(2016)\citenamefont {Kuo},
  \citenamefont {Chu}, \citenamefont {Palmstrom}, \citenamefont {Kivelson},\
  and\ \citenamefont {Fisher}}]{Kuo}%
  \BibitemOpen
  \bibfield  {author} {\bibinfo {author} {\bibfnamefont {H.-H.}\ \bibnamefont
  {Kuo}}, \bibinfo {author} {\bibfnamefont {J.-H.}\ \bibnamefont {Chu}},
  \bibinfo {author} {\bibfnamefont {J.~C.}\ \bibnamefont {Palmstrom}}, \bibinfo
  {author} {\bibfnamefont {S.~A.}\ \bibnamefont {Kivelson}},\ and\ \bibinfo
  {author} {\bibfnamefont {I.~R.}\ \bibnamefont {Fisher}},\ }\bibfield  {title}
  {\bibinfo {title} {{Ubiquitous signatures of nematic quantum criticality in
  optimally doped Fe-based superconductors}},\ }\href
  {https://doi.org/10.1126/science.aab0103} {\bibfield  {journal} {\bibinfo
  {journal} {Science}\ }\textbf {\bibinfo {volume} {352}},\ \bibinfo {pages}
  {958} (\bibinfo {year} {2016})}\BibitemShut {NoStop}%
\bibitem [{\citenamefont {Hosoi}\ \emph {et~al.}(2016)\citenamefont {Hosoi},
  \citenamefont {Matsuura}, \citenamefont {Ishida}, \citenamefont {Wang},
  \citenamefont {Mizukami}, \citenamefont {Watashige}, \citenamefont
  {Kasahara}, \citenamefont {Matsuda},\ and\ \citenamefont
  {Shibauchi}}]{Hosoi}%
  \BibitemOpen
  \bibfield  {author} {\bibinfo {author} {\bibfnamefont {S.}~\bibnamefont
  {Hosoi}}, \bibinfo {author} {\bibfnamefont {K.}~\bibnamefont {Matsuura}},
  \bibinfo {author} {\bibfnamefont {K.}~\bibnamefont {Ishida}}, \bibinfo
  {author} {\bibfnamefont {H.}~\bibnamefont {Wang}}, \bibinfo {author}
  {\bibfnamefont {Y.}~\bibnamefont {Mizukami}}, \bibinfo {author}
  {\bibfnamefont {T.}~\bibnamefont {Watashige}}, \bibinfo {author}
  {\bibfnamefont {S.}~\bibnamefont {Kasahara}}, \bibinfo {author}
  {\bibfnamefont {Y.}~\bibnamefont {Matsuda}},\ and\ \bibinfo {author}
  {\bibfnamefont {T.}~\bibnamefont {Shibauchi}},\ }\bibfield  {title} {\bibinfo
  {title} {{Nematic quantum critical point without magnetism in
  FeSe$_{1-x}$S$_x$ superconductors}},\ }\href
  {https://doi.org/10.1073/pnas.1605806113} {\bibfield  {journal} {\bibinfo
  {journal} {PNAS}\ }\textbf {\bibinfo {volume} {113}},\ \bibinfo {pages}
  {81390} (\bibinfo {year} {2016})}\BibitemShut {NoStop}%
\bibitem [{\citenamefont {Ishida}\ \emph {et~al.}(2020)\citenamefont {Ishida},
  \citenamefont {Tsujii}, \citenamefont {Hosoi}, \citenamefont {Mizukami},
  \citenamefont {Ishida}, \citenamefont {Iyo}, \citenamefont {Eisaki},
  \citenamefont {Wolf}, \citenamefont {Grube}, \citenamefont {v.~L\"ohneysen},
  \citenamefont {Fernandes},\ and\ \citenamefont {Shibauchi}}]{Ishida}%
  \BibitemOpen
  \bibfield  {author} {\bibinfo {author} {\bibfnamefont {K.}~\bibnamefont
  {Ishida}}, \bibinfo {author} {\bibfnamefont {M.}~\bibnamefont {Tsujii}},
  \bibinfo {author} {\bibfnamefont {S.}~\bibnamefont {Hosoi}}, \bibinfo
  {author} {\bibfnamefont {Y.}~\bibnamefont {Mizukami}}, \bibinfo {author}
  {\bibfnamefont {S.}~\bibnamefont {Ishida}}, \bibinfo {author} {\bibfnamefont
  {A.}~\bibnamefont {Iyo}}, \bibinfo {author} {\bibfnamefont {H.}~\bibnamefont
  {Eisaki}}, \bibinfo {author} {\bibfnamefont {T.}~\bibnamefont {Wolf}},
  \bibinfo {author} {\bibfnamefont {K.}~\bibnamefont {Grube}}, \bibinfo
  {author} {\bibfnamefont {H.}~\bibnamefont {v.~L\"ohneysen}}, \bibinfo
  {author} {\bibfnamefont {R.~M.}\ \bibnamefont {Fernandes}},\ and\ \bibinfo
  {author} {\bibfnamefont {T.}~\bibnamefont {Shibauchi}},\ }\bibfield  {title}
  {\bibinfo {title} {{Novel electronic nematicity in heavily hole-doped iron
  pnictide superconductors}},\ }\href {https://doi.org/10.1073/pnas.1909172117}
  {\bibfield  {journal} {\bibinfo  {journal} {PNAS}\ }\textbf {\bibinfo
  {volume} {117}},\ \bibinfo {pages} {6424} (\bibinfo {year}
  {2020})}\BibitemShut {NoStop}%
\bibitem [{\citenamefont {Wiecki}\ \emph {et~al.}(2020)\citenamefont {Wiecki},
  \citenamefont {Haghighirad}, \citenamefont {Weber}, \citenamefont {Merz},
  \citenamefont {Heid},\ and\ \citenamefont {B\"ohmer}}]{PRLWiecki}%
  \BibitemOpen
  \bibfield  {author} {\bibinfo {author} {\bibfnamefont {P.}~\bibnamefont
  {Wiecki}}, \bibinfo {author} {\bibfnamefont {A.-A.}\ \bibnamefont
  {Haghighirad}}, \bibinfo {author} {\bibfnamefont {F.}~\bibnamefont {Weber}},
  \bibinfo {author} {\bibfnamefont {M.}~\bibnamefont {Merz}}, \bibinfo {author}
  {\bibfnamefont {R.}~\bibnamefont {Heid}},\ and\ \bibinfo {author}
  {\bibfnamefont {A.~E.}\ \bibnamefont {B\"ohmer}},\ }\bibfield  {title}
  {\bibinfo {title} {{Dominant In-Plane Symmetric Elastoresistance in
  ${\mathrm{CsFe}}_{2}{\mathrm{As}}_{2}$}},\ }\href
  {https://doi.org/10.1103/PhysRevLett.125.187001} {\bibfield  {journal}
  {\bibinfo  {journal} {Phys. Rev. Lett.}\ }\textbf {\bibinfo {volume} {125}},\
  \bibinfo {pages} {187001} (\bibinfo {year} {2020})}\BibitemShut {NoStop}%
\bibitem [{\citenamefont {Wiecki}\ \emph {et~al.}(2021)\citenamefont {Wiecki},
  \citenamefont {Frachet}, \citenamefont {Haghighirad}, \citenamefont {Wolf},
  \citenamefont {Meingast}, \citenamefont {Heid},\ and\ \citenamefont
  {B{\"o}hmer}}]{Wiecki2021}%
  \BibitemOpen
  \bibfield  {author} {\bibinfo {author} {\bibfnamefont {P.}~\bibnamefont
  {Wiecki}}, \bibinfo {author} {\bibfnamefont {M.}~\bibnamefont {Frachet}},
  \bibinfo {author} {\bibfnamefont {A.-A.}\ \bibnamefont {Haghighirad}},
  \bibinfo {author} {\bibfnamefont {T.}~\bibnamefont {Wolf}}, \bibinfo {author}
  {\bibfnamefont {C.}~\bibnamefont {Meingast}}, \bibinfo {author}
  {\bibfnamefont {R.}~\bibnamefont {Heid}},\ and\ \bibinfo {author}
  {\bibfnamefont {A.~E.}\ \bibnamefont {B{\"o}hmer}},\ }\bibfield  {title}
  {\bibinfo {title} {Emerging symmetric strain response and weakening nematic
  fluctuations in strongly hole-doped iron-based superconductors},\ }\href
  {https://doi.org/10.1038/s41467-021-25121-5} {\bibfield  {journal} {\bibinfo
  {journal} {Nat. Commun.}\ }\textbf {\bibinfo {volume} {12}},\ \bibinfo
  {pages} {4824} (\bibinfo {year} {2021})}\BibitemShut {NoStop}%
\bibitem [{\citenamefont {Worasaran}\ \emph {et~al.}(2021)\citenamefont
  {Worasaran}, \citenamefont {Ikeda}, \citenamefont {Palmstrom}, \citenamefont
  {Straquadine}, \citenamefont {Kivelson},\ and\ \citenamefont
  {Fisher}}]{Worasaran}%
  \BibitemOpen
  \bibfield  {author} {\bibinfo {author} {\bibfnamefont {T.}~\bibnamefont
  {Worasaran}}, \bibinfo {author} {\bibfnamefont {M.~S.}\ \bibnamefont
  {Ikeda}}, \bibinfo {author} {\bibfnamefont {J.~C.}\ \bibnamefont
  {Palmstrom}}, \bibinfo {author} {\bibfnamefont {J.~A.~W.}\ \bibnamefont
  {Straquadine}}, \bibinfo {author} {\bibfnamefont {S.~A.}\ \bibnamefont
  {Kivelson}},\ and\ \bibinfo {author} {\bibfnamefont {I.~R.}\ \bibnamefont
  {Fisher}},\ }\bibfield  {title} {\bibinfo {title} {{Nematic quantum
  criticality in an Fe-based superconductor revealed by strain-tuning}},\
  }\href {https://doi.org/10.1126/science.abb9280} {\bibfield  {journal}
  {\bibinfo  {journal} {Science}\ }\textbf {\bibinfo {volume} {372}},\ \bibinfo
  {pages} {973} (\bibinfo {year} {2021})}\BibitemShut {NoStop}%
\bibitem [{\citenamefont {Rosenberg}\ \emph {et~al.}(2019)\citenamefont
  {Rosenberg}, \citenamefont {Chu}, \citenamefont {Ruff},\ and\ \citenamefont
  {Fisher}}]{Rosenberg2019}%
  \BibitemOpen
  \bibfield  {author} {\bibinfo {author} {\bibfnamefont {E.~W.}\ \bibnamefont
  {Rosenberg}}, \bibinfo {author} {\bibfnamefont {J.-H.}\ \bibnamefont {Chu}},
  \bibinfo {author} {\bibfnamefont {J.~P.~C.}\ \bibnamefont {Ruff}},\ and\
  \bibinfo {author} {\bibfnamefont {I.~R.}\ \bibnamefont {Fisher}},\ }\bibfield
   {title} {\bibinfo {title} {{Divergence of the quadrupole-strain
  susceptibility of the electronic nematic system YbRu$_2$Ge$_2$}},\ }\href
  {https://doi.org/10.1073/pnas.1818910116} {\bibfield  {journal} {\bibinfo
  {journal} {PNAS}\ }\textbf {\bibinfo {volume} {116}},\ \bibinfo {pages}
  {7232} (\bibinfo {year} {2019})}\BibitemShut {NoStop}%
\bibitem [{\citenamefont {Ye}\ \emph {et~al.}(2023)\citenamefont {Ye},
  \citenamefont {Sun}, \citenamefont {Sunko},\ and\ \citenamefont
  {Fisher}}]{Ye2023}%
  \BibitemOpen
  \bibfield  {author} {\bibinfo {author} {\bibfnamefont {L.}~\bibnamefont
  {Ye}}, \bibinfo {author} {\bibfnamefont {Y.}~\bibnamefont {Sun}}, \bibinfo
  {author} {\bibfnamefont {V.}~\bibnamefont {Sunko}},\ and\ \bibinfo {author}
  {\bibfnamefont {I.~R.}\ \bibnamefont {Fisher}},\ }\bibfield  {title}
  {\bibinfo {title} {{Elastocaloric signatures of symmetric and antisymmetric
  strain-tuning of quadrupolar and magnetic phases in DyB$_2$C$_2$}},\ }\href
  {https://doi.org/10.1073/pnas.2302800120} {\bibfield  {journal} {\bibinfo
  {journal} {PNAS}\ }\textbf {\bibinfo {volume} {120}},\ \bibinfo {pages}
  {e2302800120} (\bibinfo {year} {2023})}\BibitemShut {NoStop}%
\bibitem [{\citenamefont {Ye}\ \emph {et~al.}(2024)\citenamefont {Ye},
  \citenamefont {Sorensen}, \citenamefont {Bachmann},\ and\ \citenamefont
  {Fisher}}]{Ye2024}%
  \BibitemOpen
  \bibfield  {author} {\bibinfo {author} {\bibfnamefont {L.}~\bibnamefont
  {Ye}}, \bibinfo {author} {\bibfnamefont {M.~E.}\ \bibnamefont {Sorensen}},
  \bibinfo {author} {\bibfnamefont {M.~D.}\ \bibnamefont {Bachmann}},\ and\
  \bibinfo {author} {\bibfnamefont {I.~R.}\ \bibnamefont {Fisher}},\ }\bibfield
   {title} {\bibinfo {title} {{Measurement of the magnetic octupole
  susceptibility of PrV$_2$Al$_{20}$}},\ }\href
  {https://doi.org/10.1038/s41467-024-51269-x} {\bibfield  {journal} {\bibinfo
  {journal} {Nat. Commun.}\ }\textbf {\bibinfo {volume} {15}},\ \bibinfo
  {pages} {7005} (\bibinfo {year} {2024})}\BibitemShut {NoStop}%
\bibitem [{\citenamefont {Rosenberg}\ \emph {et~al.}(2024)\citenamefont
  {Rosenberg}, \citenamefont {Ikeda},\ and\ \citenamefont
  {Fisher}}]{Rosenberg2024}%
  \BibitemOpen
  \bibfield  {author} {\bibinfo {author} {\bibfnamefont {E.~W.}\ \bibnamefont
  {Rosenberg}}, \bibinfo {author} {\bibfnamefont {M.}~\bibnamefont {Ikeda}},\
  and\ \bibinfo {author} {\bibfnamefont {I.~R.}\ \bibnamefont {Fisher}},\
  }\bibfield  {title} {\bibinfo {title} {{The nematic susceptibility of the
  ferroquadrupolar metal TmAg$_2$ measured via the elastocaloric effect}},\
  }\href {https://doi.org/10.1038/s41535-024-00658-y} {\bibfield  {journal}
  {\bibinfo  {journal} {npj Quantum Mater.}\ }\textbf {\bibinfo {volume} {9}},\
  \bibinfo {pages} {46} (\bibinfo {year} {2024})}\BibitemShut {NoStop}%
\bibitem [{\citenamefont {Gegenwart}\ \emph {et~al.}(2008)\citenamefont
  {Gegenwart}, \citenamefont {Si},\ and\ \citenamefont
  {Steglich}}]{Gegenwart2008}%
  \BibitemOpen
  \bibfield  {author} {\bibinfo {author} {\bibfnamefont {P.}~\bibnamefont
  {Gegenwart}}, \bibinfo {author} {\bibfnamefont {Q.}~\bibnamefont {Si}},\ and\
  \bibinfo {author} {\bibfnamefont {F.}~\bibnamefont {Steglich}},\ }\bibfield
  {title} {\bibinfo {title} {Quantum criticality in heavy-fermion metals},\
  }\href {https://doi.org/10.1038/nphys892} {\bibfield  {journal} {\bibinfo
  {journal} {Nat. Phys.}\ }\textbf {\bibinfo {volume} {4}},\ \bibinfo {pages}
  {186} (\bibinfo {year} {2008})}\BibitemShut {NoStop}%
\bibitem [{\citenamefont {Stewart}(1984)}]{Stewart1984}%
  \BibitemOpen
  \bibfield  {author} {\bibinfo {author} {\bibfnamefont {G.~R.}\ \bibnamefont
  {Stewart}},\ }\bibfield  {title} {\bibinfo {title} {Heavy-fermion systems},\
  }\href {https://doi.org/10.1103/RevModPhys.56.755} {\bibfield  {journal}
  {\bibinfo  {journal} {Rev. Mod. Phys.}\ }\textbf {\bibinfo {volume} {56}},\
  \bibinfo {pages} {755} (\bibinfo {year} {1984})}\BibitemShut {NoStop}%
\bibitem [{\citenamefont {Doniach}(1977)}]{Doniach1977}%
  \BibitemOpen
  \bibfield  {author} {\bibinfo {author} {\bibfnamefont {S.}~\bibnamefont
  {Doniach}},\ }\bibfield  {title} {\bibinfo {title} {The kondo lattice and
  weak antiferromagnetism},\ }\href
  {https://doi.org/https://doi.org/10.1016/0378-4363(77)90190-5} {\bibfield
  {journal} {\bibinfo  {journal} {Physica B+C}\ }\textbf {\bibinfo {volume}
  {91}},\ \bibinfo {pages} {231} (\bibinfo {year} {1977})}\BibitemShut
  {NoStop}%
\bibitem [{\citenamefont {Gegenwart}(2016)}]{Gegenwart2016}%
  \BibitemOpen
  \bibfield  {author} {\bibinfo {author} {\bibfnamefont {P.}~\bibnamefont
  {Gegenwart}},\ }\bibfield  {title} {\bibinfo {title} {Grüneisen parameter
  studies on heavy fermion quantum criticality},\ }\href
  {https://doi.org/10.1088/0034-4885/79/11/114502} {\bibfield  {journal}
  {\bibinfo  {journal} {Rep. Prog. Phys.}\ }\textbf {\bibinfo {volume} {79}},\
  \bibinfo {pages} {114502} (\bibinfo {year} {2016})}\BibitemShut {NoStop}%
\bibitem [{\citenamefont {Lavagna}\ \emph {et~al.}(1982)\citenamefont
  {Lavagna}, \citenamefont {Lacroix},\ and\ \citenamefont
  {Cyrot}}]{Lavagna1982}%
  \BibitemOpen
  \bibfield  {author} {\bibinfo {author} {\bibfnamefont {M.}~\bibnamefont
  {Lavagna}}, \bibinfo {author} {\bibfnamefont {C.}~\bibnamefont {Lacroix}},\
  and\ \bibinfo {author} {\bibfnamefont {M.}~\bibnamefont {Cyrot}},\ }\bibfield
   {title} {\bibinfo {title} {Electrical resistivity of the kondo lattice},\
  }\href {https://doi.org/10.1063/1.330742} {\bibfield  {journal} {\bibinfo
  {journal} {J. Appl. Phys.}\ }\textbf {\bibinfo {volume} {53}},\ \bibinfo
  {pages} {2055} (\bibinfo {year} {1982})}\BibitemShut {NoStop}%
\bibitem [{\citenamefont {Coleman}(1987)}]{Coleman1987}%
  \BibitemOpen
  \bibfield  {author} {\bibinfo {author} {\bibfnamefont {P.}~\bibnamefont
  {Coleman}},\ }\bibfield  {title} {\bibinfo {title} {Constrained
  quasiparticles and conduction in heavy-fermion systems},\ }\href
  {https://doi.org/10.1103/PhysRevLett.59.1026} {\bibfield  {journal} {\bibinfo
   {journal} {Phys. Rev. Lett.}\ }\textbf {\bibinfo {volume} {59}},\ \bibinfo
  {pages} {10265} (\bibinfo {year} {1987})}\BibitemShut {NoStop}%
\bibitem [{\citenamefont {Cox}\ and\ \citenamefont {Grewe}(1988)}]{Cox1988}%
  \BibitemOpen
  \bibfield  {author} {\bibinfo {author} {\bibfnamefont {D.~L.}\ \bibnamefont
  {Cox}}\ and\ \bibinfo {author} {\bibfnamefont {N.}~\bibnamefont {Grewe}},\
  }\bibfield  {title} {\bibinfo {title} {Transport properties of the anderson
  lattice},\ }\href {https://doi.org/10.1007/BF01312492} {\bibfield  {journal}
  {\bibinfo  {journal} {Z. Phys. B}\ }\textbf {\bibinfo {volume} {71}},\
  \bibinfo {pages} {321} (\bibinfo {year} {1988})}\BibitemShut {NoStop}%
\bibitem [{\citenamefont {Schlottmann}(1983)}]{Schlottmann1983}%
  \BibitemOpen
  \bibfield  {author} {\bibinfo {author} {\bibfnamefont {P.}~\bibnamefont
  {Schlottmann}},\ }\bibfield  {title} {\bibinfo {title} {{Bethe-Ansatz
  Solution of the Ground-State of the SU(2j+ 1) Kondo(Coqblin-Schrieffer)
  Model: Magnetization, Magnetoresistance and Universality}},\ }\href
  {https://doi.org/10.1007/BF01307678} {\bibfield  {journal} {\bibinfo
  {journal} {Z. Phys. B}\ }\textbf {\bibinfo {volume} {51}},\ \bibinfo {pages}
  {223} (\bibinfo {year} {1983})}\BibitemShut {NoStop}%
\bibitem [{\citenamefont {Batlogg}\ \emph {et~al.}(1987)\citenamefont
  {Batlogg}, \citenamefont {Bishop}, \citenamefont {Bucher}, \citenamefont
  {Golding}, \citenamefont {Ramirez}, \citenamefont {Fisk}, \citenamefont
  {Smith},\ and\ \citenamefont {Ott}}]{Batlogg1987}%
  \BibitemOpen
  \bibfield  {author} {\bibinfo {author} {\bibfnamefont {B.}~\bibnamefont
  {Batlogg}}, \bibinfo {author} {\bibfnamefont {D.}~\bibnamefont {Bishop}},
  \bibinfo {author} {\bibfnamefont {E.}~\bibnamefont {Bucher}}, \bibinfo
  {author} {\bibfnamefont {B.}~\bibnamefont {Golding}}, \bibinfo {author}
  {\bibfnamefont {A.}~\bibnamefont {Ramirez}}, \bibinfo {author} {\bibfnamefont
  {Z.}~\bibnamefont {Fisk}}, \bibinfo {author} {\bibfnamefont {J.}~\bibnamefont
  {Smith}},\ and\ \bibinfo {author} {\bibfnamefont {H.}~\bibnamefont {Ott}},\
  }\bibfield  {title} {\bibinfo {title} {{Superconductivity and Heavy
  Fermions}},\ }\href {https://doi.org/10.1016/0304-8853(87)90632-99}
  {\bibfield  {journal} {\bibinfo  {journal} {J. Magn. Magn. Mater.}\ }\textbf
  {\bibinfo {volume} {63}},\ \bibinfo {pages} {441} (\bibinfo {year}
  {1987})}\BibitemShut {NoStop}%
\bibitem [{\citenamefont {Gegenwart}\ \emph {et~al.}(2002)\citenamefont
  {Gegenwart}, \citenamefont {Custers}, \citenamefont {Geibel}, \citenamefont
  {Neumaier}, \citenamefont {Tayama}, \citenamefont {Tenya}, \citenamefont
  {Trovarelli},\ and\ \citenamefont {Steglich}}]{Gegenwart2002}%
  \BibitemOpen
  \bibfield  {author} {\bibinfo {author} {\bibfnamefont {P.}~\bibnamefont
  {Gegenwart}}, \bibinfo {author} {\bibfnamefont {J.}~\bibnamefont {Custers}},
  \bibinfo {author} {\bibfnamefont {C.}~\bibnamefont {Geibel}}, \bibinfo
  {author} {\bibfnamefont {K.}~\bibnamefont {Neumaier}}, \bibinfo {author}
  {\bibfnamefont {T.}~\bibnamefont {Tayama}}, \bibinfo {author} {\bibfnamefont
  {K.}~\bibnamefont {Tenya}}, \bibinfo {author} {\bibfnamefont
  {O.}~\bibnamefont {Trovarelli}},\ and\ \bibinfo {author} {\bibfnamefont
  {F.}~\bibnamefont {Steglich}},\ }\bibfield  {title} {\bibinfo {title}
  {{Magnetic-Field Induced Quantum Critical Point in
  $\mathrm{Y}\mathrm{b}\mathrm{R}{\mathrm{h}}_{\mathrm{2}}\mathrm{S}{\mathrm{i}}_{\mathrm{2}}$}},\
  }\href {https://doi.org/10.1103/PhysRevLett.89.056402} {\bibfield  {journal}
  {\bibinfo  {journal} {Phys. Rev. Lett.}\ }\textbf {\bibinfo {volume} {89}},\
  \bibinfo {pages} {056402} (\bibinfo {year} {2002})}\BibitemShut {NoStop}%
\bibitem [{\citenamefont {Tokiwa}\ \emph {et~al.}(2005)\citenamefont {Tokiwa},
  \citenamefont {Gegenwart}, \citenamefont {Radu}, \citenamefont {Ferstl},
  \citenamefont {Sparn}, \citenamefont {Geibel},\ and\ \citenamefont
  {Steglich}}]{PRLTokiwa}%
  \BibitemOpen
  \bibfield  {author} {\bibinfo {author} {\bibfnamefont {Y.}~\bibnamefont
  {Tokiwa}}, \bibinfo {author} {\bibfnamefont {P.}~\bibnamefont {Gegenwart}},
  \bibinfo {author} {\bibfnamefont {T.}~\bibnamefont {Radu}}, \bibinfo {author}
  {\bibfnamefont {J.}~\bibnamefont {Ferstl}}, \bibinfo {author} {\bibfnamefont
  {G.}~\bibnamefont {Sparn}}, \bibinfo {author} {\bibfnamefont
  {C.}~\bibnamefont {Geibel}},\ and\ \bibinfo {author} {\bibfnamefont
  {F.}~\bibnamefont {Steglich}},\ }\bibfield  {title} {\bibinfo {title}
  {{Field-Induced Suppression of the Heavy-Fermion State in
  $\mathrm{Y}\mathrm{b}\mathrm{R}{\mathrm{h}}_{\mathrm{2}}\mathrm{S}{\mathrm{i}}_{\mathrm{2}}$}},\
  }\href {https://doi.org/10.1103/PhysRevLett.94.226402} {\bibfield  {journal}
  {\bibinfo  {journal} {Phys. Rev. Lett.}\ }\textbf {\bibinfo {volume} {94}},\
  \bibinfo {pages} {226402} (\bibinfo {year} {2005})}\BibitemShut {NoStop}%
\bibitem [{\citenamefont {Custers}\ \emph {et~al.}(2003)\citenamefont
  {Custers}, \citenamefont {Gegenwart}, \citenamefont {Wilhelm}, \citenamefont
  {Neumaier}, \citenamefont {Tokiwa}, \citenamefont {Trovarelli}, \citenamefont
  {Geibel}, \citenamefont {Steglich}, \citenamefont {P{\'e}pin},\ and\
  \citenamefont {Coleman}}]{Custers2003}%
  \BibitemOpen
  \bibfield  {author} {\bibinfo {author} {\bibfnamefont {J.}~\bibnamefont
  {Custers}}, \bibinfo {author} {\bibfnamefont {P.}~\bibnamefont {Gegenwart}},
  \bibinfo {author} {\bibfnamefont {H.}~\bibnamefont {Wilhelm}}, \bibinfo
  {author} {\bibfnamefont {K.}~\bibnamefont {Neumaier}}, \bibinfo {author}
  {\bibfnamefont {Y.}~\bibnamefont {Tokiwa}}, \bibinfo {author} {\bibfnamefont
  {O.}~\bibnamefont {Trovarelli}}, \bibinfo {author} {\bibfnamefont
  {C.}~\bibnamefont {Geibel}}, \bibinfo {author} {\bibfnamefont
  {F.}~\bibnamefont {Steglich}}, \bibinfo {author} {\bibfnamefont
  {C.}~\bibnamefont {P{\'e}pin}},\ and\ \bibinfo {author} {\bibfnamefont
  {P.}~\bibnamefont {Coleman}},\ }\bibfield  {title} {\bibinfo {title} {The
  break-up of heavy electrons at a quantum critical point},\ }\href
  {https://doi.org/10.1038/nature01774} {\bibfield  {journal} {\bibinfo
  {journal} {Nature}\ }\textbf {\bibinfo {volume} {424}},\ \bibinfo {pages}
  {524} (\bibinfo {year} {2003})}\BibitemShut {NoStop}%
\bibitem [{\citenamefont {Gegenwart}\ \emph {et~al.}(2005)\citenamefont
  {Gegenwart}, \citenamefont {Custers}, \citenamefont {Tokiwa}, \citenamefont
  {Geibel},\ and\ \citenamefont {Steglich}}]{Gegenwart2005}%
  \BibitemOpen
  \bibfield  {author} {\bibinfo {author} {\bibfnamefont {P.}~\bibnamefont
  {Gegenwart}}, \bibinfo {author} {\bibfnamefont {J.}~\bibnamefont {Custers}},
  \bibinfo {author} {\bibfnamefont {Y.}~\bibnamefont {Tokiwa}}, \bibinfo
  {author} {\bibfnamefont {C.}~\bibnamefont {Geibel}},\ and\ \bibinfo {author}
  {\bibfnamefont {F.}~\bibnamefont {Steglich}},\ }\bibfield  {title} {\bibinfo
  {title} {{Ferromagnetic Quantum Critical Fluctuations in
  YbRh$_2$(Si$_{0.95}$Ge$_{0.05}$)$_2$}},\ }\href
  {https://doi.org/10.1103/PhysRevLett.94.076402} {\bibfield  {journal}
  {\bibinfo  {journal} {Phys. Rev. Lett.}\ }\textbf {\bibinfo {volume} {94}},\
  \bibinfo {pages} {076402} (\bibinfo {year} {2005})}\BibitemShut {NoStop}%
\bibitem [{\citenamefont {Panja}\ \emph {et~al.}(2024)\citenamefont {Panja},
  \citenamefont {Jesche}, \citenamefont {Tang},\ and\ \citenamefont
  {Gegenwart}}]{Panja2024}%
  \BibitemOpen
  \bibfield  {author} {\bibinfo {author} {\bibfnamefont {S.~N.}\ \bibnamefont
  {Panja}}, \bibinfo {author} {\bibfnamefont {A.}~\bibnamefont {Jesche}},
  \bibinfo {author} {\bibfnamefont {N.}~\bibnamefont {Tang}},\ and\ \bibinfo
  {author} {\bibfnamefont {P.}~\bibnamefont {Gegenwart}},\ }\bibfield  {title}
  {\bibinfo {title} {Tensile and compressive strain tuning of a kondo
  lattice},\ }\href {https://doi.org/10.1103/PhysRevB.109.205152} {\bibfield
  {journal} {\bibinfo  {journal} {Phys. Rev. B}\ }\textbf {\bibinfo {volume}
  {109}},\ \bibinfo {pages} {205152} (\bibinfo {year} {2024})}\BibitemShut
  {NoStop}%
\bibitem [{\citenamefont {Hicks}\ \emph {et~al.}(2014)\citenamefont {Hicks},
  \citenamefont {Barber}, \citenamefont {Edkins}, \citenamefont {Brodsky},\
  and\ \citenamefont {Mackenzie}}]{Hicks2014Rsi}%
  \BibitemOpen
  \bibfield  {author} {\bibinfo {author} {\bibfnamefont {C.~W.}\ \bibnamefont
  {Hicks}}, \bibinfo {author} {\bibfnamefont {M.~E.}\ \bibnamefont {Barber}},
  \bibinfo {author} {\bibfnamefont {S.~D.}\ \bibnamefont {Edkins}}, \bibinfo
  {author} {\bibfnamefont {D.~O.}\ \bibnamefont {Brodsky}},\ and\ \bibinfo
  {author} {\bibfnamefont {A.~P.}\ \bibnamefont {Mackenzie}},\ }\bibfield
  {title} {\bibinfo {title} {Piezoelectric-based apparatus for strain tuning},\
  }\href {https://doi.org/10.1063/1.4881611} {\bibfield  {journal} {\bibinfo
  {journal} {Rev. Sci. Instrum.}\ }\textbf {\bibinfo {volume} {85}},\ \bibinfo
  {pages} {065003} (\bibinfo {year} {2014})}\BibitemShut {NoStop}%
\bibitem [{sup()}]{supple}%
  \BibitemOpen
  \href@noop {} {\bibinfo {title} {See supplementary material}},\ \bibinfo
  {note} {for additional experimental details and data analysis including
  Refs.~\cite{deJong2015,Hicks2014Rsi,Kuchler2012,Panja2025,PRLWiecki,Fisher1968,Kuechler2004,Kuechler2003}.}\BibitemShut
  {Stop}%
\bibitem [{\citenamefont {Callister}\ and\ \citenamefont
  {Rethwisch}(2018)}]{Callister2018}%
  \BibitemOpen
  \bibfield  {author} {\bibinfo {author} {\bibfnamefont {W.~D.}\ \bibnamefont
  {Callister}}\ and\ \bibinfo {author} {\bibfnamefont {D.~G.}\ \bibnamefont
  {Rethwisch}},\ }\href@noop {} {\emph {\bibinfo {title} {Materials Science and
  Engineering: An Introduction}}},\ \bibinfo {edition} {10th}\ ed.\ (\bibinfo
  {publisher} {Wiley},\ \bibinfo {address} {Hoboken, NJ},\ \bibinfo {year}
  {2018})\BibitemShut {NoStop}%
\bibitem [{\citenamefont {Sun}\ \emph {et~al.}(2010)\citenamefont {Sun},
  \citenamefont {Thompson},\ and\ \citenamefont {Nishida}}]{Sun2010Strain}%
  \BibitemOpen
  \bibfield  {author} {\bibinfo {author} {\bibfnamefont {Y.}~\bibnamefont
  {Sun}}, \bibinfo {author} {\bibfnamefont {S.}~\bibnamefont {Thompson}},\ and\
  \bibinfo {author} {\bibfnamefont {T.}~\bibnamefont {Nishida}},\ }\href@noop
  {} {\emph {\bibinfo {title} {Strain Effect in Semiconductors: Theory and
  Device Applications}}}\ (\bibinfo  {publisher} {Springer},\ \bibinfo
  {address} {New York},\ \bibinfo {year} {2010})\BibitemShut {NoStop}%
\bibitem [{\citenamefont {Fisher}\ and\ \citenamefont
  {Langer}(1968)}]{Fisher1968}%
  \BibitemOpen
  \bibfield  {author} {\bibinfo {author} {\bibfnamefont {M.~E.}\ \bibnamefont
  {Fisher}}\ and\ \bibinfo {author} {\bibfnamefont {J.~S.}\ \bibnamefont
  {Langer}},\ }\bibfield  {title} {\bibinfo {title} {Resistive anomalies at
  magnetic critical points},\ }\href
  {https://doi.org/10.1103/PhysRevLett.20.665} {\bibfield  {journal} {\bibinfo
  {journal} {Phys. Rev. Lett.}\ }\textbf {\bibinfo {volume} {20}},\ \bibinfo
  {pages} {665} (\bibinfo {year} {1968})}\BibitemShut {NoStop}%
\bibitem [{\citenamefont {Küchler}\ \emph {et~al.}(2026)\citenamefont
  {Küchler}, \citenamefont {Panja}, \citenamefont {Wirth},\ and\ \citenamefont
  {Gegenwart}}]{Panja2025}%
  \BibitemOpen
  \bibfield  {author} {\bibinfo {author} {\bibfnamefont {R.}~\bibnamefont
  {Küchler}}, \bibinfo {author} {\bibfnamefont {S.}~\bibnamefont {Panja}},
  \bibinfo {author} {\bibfnamefont {S.}~\bibnamefont {Wirth}},\ and\ \bibinfo
  {author} {\bibfnamefont {P.}~\bibnamefont {Gegenwart}},\ }\bibfield  {title}
  {\bibinfo {title} {High-resolution capacitance dilatometry of microscopically
  thin samples using the world’s smallest dilatometer},\ }\href
  {https://doi.org/10.1063/5.0300507} {\bibfield  {journal} {\bibinfo
  {journal} {Rev. Sci. Instr.}\ }\textbf {\bibinfo {volume} {97}},\ \bibinfo
  {pages} {025209} (\bibinfo {year} {2026})}\BibitemShut {NoStop}%
\bibitem [{\citenamefont {K\"uchler}\ \emph {et~al.}(2004)\citenamefont
  {K\"uchler}, \citenamefont {Weickert}, \citenamefont {Gegenwart},
  \citenamefont {Oeschler}, \citenamefont {Ferstl}, \citenamefont {Geibel},\
  and\ \citenamefont {Steglich}}]{Kuechler2004}%
  \BibitemOpen
  \bibfield  {author} {\bibinfo {author} {\bibfnamefont {R.}~\bibnamefont
  {K\"uchler}}, \bibinfo {author} {\bibfnamefont {F.}~\bibnamefont {Weickert}},
  \bibinfo {author} {\bibfnamefont {P.}~\bibnamefont {Gegenwart}}, \bibinfo
  {author} {\bibfnamefont {N.}~\bibnamefont {Oeschler}}, \bibinfo {author}
  {\bibfnamefont {J.}~\bibnamefont {Ferstl}}, \bibinfo {author} {\bibfnamefont
  {C.}~\bibnamefont {Geibel}},\ and\ \bibinfo {author} {\bibfnamefont
  {F.}~\bibnamefont {Steglich}},\ }\bibfield  {title} {\bibinfo {title}
  {{Low-temperature thermal expansion and magnetostriction of
  YbRh$_2$(Si$_{1-x}$Ge$_{x}$)$_2$ ($x=0$ and $0.05$)}},\ }\href
  {https://doi.org/10.1016/j.jmmm.2003.11.094} {\bibfield  {journal} {\bibinfo
  {journal} {J. Magn. Magn. Mater.}\ }\textbf {\bibinfo {volume} {272-276}},\
  \bibinfo {pages} {229} (\bibinfo {year} {2004})}\BibitemShut {NoStop}%
\bibitem [{\citenamefont {K\"uchler}\ \emph {et~al.}(2003)\citenamefont
  {K\"uchler}, \citenamefont {Oeschler}, \citenamefont {Gegenwart},
  \citenamefont {Cichorek}, \citenamefont {Neumaier}, \citenamefont {Tegus},
  \citenamefont {Geibel}, \citenamefont {Mydosh}, \citenamefont {Steglich},
  \citenamefont {Zhu},\ and\ \citenamefont {Si}}]{Kuechler2003}%
  \BibitemOpen
  \bibfield  {author} {\bibinfo {author} {\bibfnamefont {R.}~\bibnamefont
  {K\"uchler}}, \bibinfo {author} {\bibfnamefont {N.}~\bibnamefont {Oeschler}},
  \bibinfo {author} {\bibfnamefont {P.}~\bibnamefont {Gegenwart}}, \bibinfo
  {author} {\bibfnamefont {T.}~\bibnamefont {Cichorek}}, \bibinfo {author}
  {\bibfnamefont {K.}~\bibnamefont {Neumaier}}, \bibinfo {author}
  {\bibfnamefont {O.}~\bibnamefont {Tegus}}, \bibinfo {author} {\bibfnamefont
  {C.}~\bibnamefont {Geibel}}, \bibinfo {author} {\bibfnamefont
  {J.}~\bibnamefont {Mydosh}}, \bibinfo {author} {\bibfnamefont
  {F.}~\bibnamefont {Steglich}}, \bibinfo {author} {\bibfnamefont
  {L.}~\bibnamefont {Zhu}},\ and\ \bibinfo {author} {\bibfnamefont
  {Q.}~\bibnamefont {Si}},\ }\bibfield  {title} {\bibinfo {title} {{Divergence
  of the Gr\"uneisen Ratio at Quantum Critical Points in Heavy Fermion
  Metals}},\ }\href {https://doi.org/10.1103/PhysRevLett.91.066405} {\bibfield
  {journal} {\bibinfo  {journal} {Phys. Rev. Lett.}\ }\textbf {\bibinfo
  {volume} {91}},\ \bibinfo {pages} {066405} (\bibinfo {year}
  {2003})}\BibitemShut {NoStop}%
\bibitem [{\citenamefont {Zhu}\ \emph {et~al.}(2003)\citenamefont {Zhu},
  \citenamefont {Garst}, \citenamefont {Rosch},\ and\ \citenamefont
  {Si}}]{Zhu2003}%
  \BibitemOpen
  \bibfield  {author} {\bibinfo {author} {\bibfnamefont {L.}~\bibnamefont
  {Zhu}}, \bibinfo {author} {\bibfnamefont {M.}~\bibnamefont {Garst}}, \bibinfo
  {author} {\bibfnamefont {A.}~\bibnamefont {Rosch}},\ and\ \bibinfo {author}
  {\bibfnamefont {Q.}~\bibnamefont {Si}},\ }\bibfield  {title} {\bibinfo
  {title} {{Universally Diverging Gr\"uneisen Parameter and the Magnetocaloric
  Effect Close to Quantum Critical Points}},\ }\href
  {https://doi.org/10.1103/PhysRevLett.91.066404} {\bibfield  {journal}
  {\bibinfo  {journal} {Phys. Rev. Lett.}\ }\textbf {\bibinfo {volume} {91}},\
  \bibinfo {pages} {066404} (\bibinfo {year} {2003})}\BibitemShut {NoStop}%
\bibitem [{\citenamefont {Tokiwa}\ and\ \citenamefont
  {Gegenwart}(2011)}]{Tokiwa}%
  \BibitemOpen
  \bibfield  {author} {\bibinfo {author} {\bibfnamefont {Y.}~\bibnamefont
  {Tokiwa}}\ and\ \bibinfo {author} {\bibfnamefont {P.}~\bibnamefont
  {Gegenwart}},\ }\bibfield  {title} {\bibinfo {title} {{High-resolution
  alternating-field technique to determine the magnetocaloric effect of metals
  down to very low temperatures}},\ }\href {https://doi.org/10.1063/1.3529433}
  {\bibfield  {journal} {\bibinfo  {journal} {Rev. Sci. Instrum.}\ }\textbf
  {\bibinfo {volume} {82}},\ \bibinfo {pages} {013905} (\bibinfo {year}
  {2011})}\BibitemShut {NoStop}%
\bibitem [{\citenamefont {Li}\ \emph {et~al.}(2022)\citenamefont {Li},
  \citenamefont {Garst}, \citenamefont {Schmalian}, \citenamefont {Ghosh},
  \citenamefont {Kikugawa}, \citenamefont {Sokolov}, \citenamefont {Hicks},
  \citenamefont {Jerzembeck}, \citenamefont {Ikeda}, \citenamefont {Hu},
  \citenamefont {Ramshaw}, \citenamefont {Rost}, \citenamefont {Nicklas},\ and\
  \citenamefont {Mackenzie}}]{Li2022}%
  \BibitemOpen
  \bibfield  {author} {\bibinfo {author} {\bibfnamefont {Y.-S.}\ \bibnamefont
  {Li}}, \bibinfo {author} {\bibfnamefont {M.}~\bibnamefont {Garst}}, \bibinfo
  {author} {\bibfnamefont {J.}~\bibnamefont {Schmalian}}, \bibinfo {author}
  {\bibfnamefont {S.}~\bibnamefont {Ghosh}}, \bibinfo {author} {\bibfnamefont
  {N.}~\bibnamefont {Kikugawa}}, \bibinfo {author} {\bibfnamefont
  {D.}~\bibnamefont {Sokolov}}, \bibinfo {author} {\bibfnamefont
  {C.}~\bibnamefont {Hicks}}, \bibinfo {author} {\bibfnamefont
  {F.}~\bibnamefont {Jerzembeck}}, \bibinfo {author} {\bibfnamefont
  {M.}~\bibnamefont {Ikeda}}, \bibinfo {author} {\bibfnamefont
  {Z.}~\bibnamefont {Hu}}, \bibinfo {author} {\bibfnamefont {B.}~\bibnamefont
  {Ramshaw}}, \bibinfo {author} {\bibfnamefont {A.}~\bibnamefont {Rost}},
  \bibinfo {author} {\bibfnamefont {M.}~\bibnamefont {Nicklas}},\ and\ \bibinfo
  {author} {\bibfnamefont {A.}~\bibnamefont {Mackenzie}},\ }\bibfield  {title}
  {\bibinfo {title} {{Elastocaloric determination of the phase diagram of
  Sr$_2$RuO$_4$}},\ }\href {https://doi.org/10.1038/s41586-022-04820-z}
  {\bibfield  {journal} {\bibinfo  {journal} {Nature}\ }\textbf {\bibinfo
  {volume} {607}},\ \bibinfo {pages} {276} (\bibinfo {year}
  {2022})}\BibitemShut {NoStop}%
\bibitem [{\citenamefont {Hlubina}\ and\ \citenamefont
  {Rice}(1995)}]{Hlubina1995}%
  \BibitemOpen
  \bibfield  {author} {\bibinfo {author} {\bibfnamefont {R.}~\bibnamefont
  {Hlubina}}\ and\ \bibinfo {author} {\bibfnamefont {T.}~\bibnamefont {Rice}},\
  }\bibfield  {title} {\bibinfo {title} {{Resistivity as a function of
  temperature for models with hot spots on the Fermi surface}},\ }\href
  {https://doi.org/10.1103/PhysRevB.51.9253} {\bibfield  {journal} {\bibinfo
  {journal} {Phys. Rev. B}\ }\textbf {\bibinfo {volume} {51}},\ \bibinfo
  {pages} {9253} (\bibinfo {year} {1995})}\BibitemShut {NoStop}%
\bibitem [{\citenamefont {Coleman}\ \emph {et~al.}(2001)\citenamefont
  {Coleman}, \citenamefont {P\'{e}in}, \citenamefont {Si},\ and\ \citenamefont
  {Ramazashvili}}]{Coleman2001}%
  \BibitemOpen
  \bibfield  {author} {\bibinfo {author} {\bibfnamefont {P.}~\bibnamefont
  {Coleman}}, \bibinfo {author} {\bibfnamefont {C.}~\bibnamefont {P\'{e}in}},
  \bibinfo {author} {\bibfnamefont {Q.}~\bibnamefont {Si}},\ and\ \bibinfo
  {author} {\bibfnamefont {R.}~\bibnamefont {Ramazashvili}},\ }\bibfield
  {title} {\bibinfo {title} {{HowdoFermi liquids get heavy and die?}},\ }\href
  {https://doi.org/10.1088/0953-8984/13/35/202} {\bibfield  {journal} {\bibinfo
   {journal} {J. Phys.: Condens. Matter}\ }\textbf {\bibinfo {volume} {13}},\
  \bibinfo {pages} {R723} (\bibinfo {year} {2001})}\BibitemShut {NoStop}%
\bibitem [{\citenamefont {Pustogow}\ \emph {et~al.}(2023)\citenamefont
  {Pustogow}, \citenamefont {Kawasugi}, \citenamefont {Sakurakoji},\ and\
  \citenamefont {Tajima}}]{Pustogow2023}%
  \BibitemOpen
  \bibfield  {author} {\bibinfo {author} {\bibfnamefont {A.}~\bibnamefont
  {Pustogow}}, \bibinfo {author} {\bibfnamefont {Y.}~\bibnamefont {Kawasugi}},
  \bibinfo {author} {\bibfnamefont {H.}~\bibnamefont {Sakurakoji}},\ and\
  \bibinfo {author} {\bibfnamefont {N.}~\bibnamefont {Tajima}},\ }\bibfield
  {title} {\bibinfo {title} {{Chasing the spin gap through the phase diagram of
  a frustrated Mott insulator}},\ }\href
  {https://doi.org/10.1038/s41467-023-37491-z} {\bibfield  {journal} {\bibinfo
  {journal} {Nat. Commun.}\ }\textbf {\bibinfo {volume} {14}},\ \bibinfo
  {pages} {1960} (\bibinfo {year} {2023})}\BibitemShut {NoStop}%
\bibitem [{\citenamefont {Schubert}\ \emph {et~al.}(2019)\citenamefont
  {Schubert}, \citenamefont {Tokiwa}, \citenamefont {H\"ubner}, \citenamefont
  {Mchalwat}, \citenamefont {Blumenr\"other}, \citenamefont {Jeevan},\ and\
  \citenamefont {Gegenwart}}]{Schubert}%
  \BibitemOpen
  \bibfield  {author} {\bibinfo {author} {\bibfnamefont {M.-H.}\ \bibnamefont
  {Schubert}}, \bibinfo {author} {\bibfnamefont {Y.}~\bibnamefont {Tokiwa}},
  \bibinfo {author} {\bibfnamefont {S.-H.}\ \bibnamefont {H\"ubner}}, \bibinfo
  {author} {\bibfnamefont {M.}~\bibnamefont {Mchalwat}}, \bibinfo {author}
  {\bibfnamefont {E.}~\bibnamefont {Blumenr\"other}}, \bibinfo {author}
  {\bibfnamefont {H.~S.}\ \bibnamefont {Jeevan}},\ and\ \bibinfo {author}
  {\bibfnamefont {P.}~\bibnamefont {Gegenwart}},\ }\bibfield  {title} {\bibinfo
  {title} {{Tuning low-energy scales in
  $\mathrm{Y}\mathrm{b}\mathrm{R}{\mathrm{h}}_{\mathrm{2}}\mathrm{S}{\mathrm{i}}_{\mathrm{2}}$
  by non-isoelectronic substitution and pressure}},\ }\href
  {https://doi.org/10.1103/PhysRevResearch.1.032004} {\bibfield  {journal}
  {\bibinfo  {journal} {Phys. Rev. Res.}\ }\textbf {\bibinfo {volume} {1}},\
  \bibinfo {pages} {032004(R)} (\bibinfo {year} {2019})}\BibitemShut {NoStop}%
\bibitem [{\citenamefont {de~Jong}\ \emph {et~al.}(2015)\citenamefont
  {de~Jong}, \citenamefont {Chen}, \citenamefont {Angsten}, \citenamefont
  {Jain}, \citenamefont {Notestine}, \citenamefont {Gamst}, \citenamefont
  {Sluiter}, \citenamefont {Krishna~Ande}, \citenamefont {van~der Zwaag},
  \citenamefont {Plata}, \citenamefont {Toher}, \citenamefont {Curtarolo},
  \citenamefont {Ceder}, \citenamefont {Persson},\ and\ \citenamefont
  {Asta}}]{deJong2015}%
  \BibitemOpen
  \bibfield  {author} {\bibinfo {author} {\bibfnamefont {M.}~\bibnamefont
  {de~Jong}}, \bibinfo {author} {\bibfnamefont {W.}~\bibnamefont {Chen}},
  \bibinfo {author} {\bibfnamefont {T.}~\bibnamefont {Angsten}}, \bibinfo
  {author} {\bibfnamefont {A.}~\bibnamefont {Jain}}, \bibinfo {author}
  {\bibfnamefont {R.}~\bibnamefont {Notestine}}, \bibinfo {author}
  {\bibfnamefont {A.}~\bibnamefont {Gamst}}, \bibinfo {author} {\bibfnamefont
  {M.}~\bibnamefont {Sluiter}}, \bibinfo {author} {\bibfnamefont
  {C.}~\bibnamefont {Krishna~Ande}}, \bibinfo {author} {\bibfnamefont
  {S.}~\bibnamefont {van~der Zwaag}}, \bibinfo {author} {\bibfnamefont {J.~J.}\
  \bibnamefont {Plata}}, \bibinfo {author} {\bibfnamefont {C.}~\bibnamefont
  {Toher}}, \bibinfo {author} {\bibfnamefont {S.}~\bibnamefont {Curtarolo}},
  \bibinfo {author} {\bibfnamefont {G.}~\bibnamefont {Ceder}}, \bibinfo
  {author} {\bibfnamefont {K.~A.}\ \bibnamefont {Persson}},\ and\ \bibinfo
  {author} {\bibfnamefont {M.}~\bibnamefont {Asta}},\ }\bibfield  {title}
  {\bibinfo {title} {Charting the complete elastic properties of inorganic
  crystalline compounds},\ }\href {https://doi.org/10.1038/sdata.2015.9}
  {\bibfield  {journal} {\bibinfo  {journal} {Sci. Data}\ }\textbf {\bibinfo
  {volume} {2}},\ \bibinfo {pages} {150009} (\bibinfo {year}
  {2015})}\BibitemShut {NoStop}%
\bibitem [{\citenamefont {Küchler}\ \emph {et~al.}(2012)\citenamefont
  {Küchler}, \citenamefont {Bauer}, \citenamefont {Brando},\ and\
  \citenamefont {Steglich}}]{Kuchler2012}%
  \BibitemOpen
  \bibfield  {author} {\bibinfo {author} {\bibfnamefont {R.}~\bibnamefont
  {Küchler}}, \bibinfo {author} {\bibfnamefont {T.}~\bibnamefont {Bauer}},
  \bibinfo {author} {\bibfnamefont {M.}~\bibnamefont {Brando}},\ and\ \bibinfo
  {author} {\bibfnamefont {F.}~\bibnamefont {Steglich}},\ }\bibfield  {title}
  {\bibinfo {title} {A compact and miniaturized high resolution capacitance
  dilatometer for measuring thermal expansion and magnetostriction},\ }\href
  {https://doi.org/10.1063/1.4748864} {\bibfield  {journal} {\bibinfo
  {journal} {Rev. Sci. Instrum.}\ }\textbf {\bibinfo {volume} {83}},\ \bibinfo
  {pages} {095102} (\bibinfo {year} {2012})}\BibitemShut {NoStop}%
\end{thebibliography}%

\newpage
\setcounter{figure}{0}
\renewcommand{\thefigure}{S\arabic{figure}} 
\renewcommand{\theequation}{S\arabic{equation}}
\setcounter{equation}{0}
\onecolumngrid

\section*{Supplemental Material}
\author{Soumendra Nath Panja}
	\author{Jacques G Pontanel}
	\author{Julian Kaiser}
	\author{Anton Jesche}
	\author{Philipp Gegenwart}
	
\maketitle

  \section{Experimental Details}
  
	  Eight electrical contacts were put on the bar- shaped, cut and polished crystal. Four contacts were aligned with the applied strain direction to measure the longitudinal elastoresistance, and four more contacts were oriented perpendicular to the strain direction to measure the transverse response, as shown in Fig. \ref{figure1}. Uniaxial strain is applied along the horizontal direction $\varepsilon_{xx}$ ($\varepsilon \parallel$ 100/110 ), and the resulting transverse strain $\varepsilon_{xx}$ ($\varepsilon \parallel$ 010/$1\bar{1}0$) follows the Poisson relation $\varepsilon_{yy} = -\nu\,\varepsilon_{xx}$.
	 This eight-contact configuration allows simultaneous measurement of the longitudinal and transverse strain responses, enabling extraction of the elastoresistivity in the $A_{1g}$, $B_{1g}$, and $B_{2g}$ symmetry channels.
	 
	\begin{figure}[h]
	\includegraphics[scale=0.50]{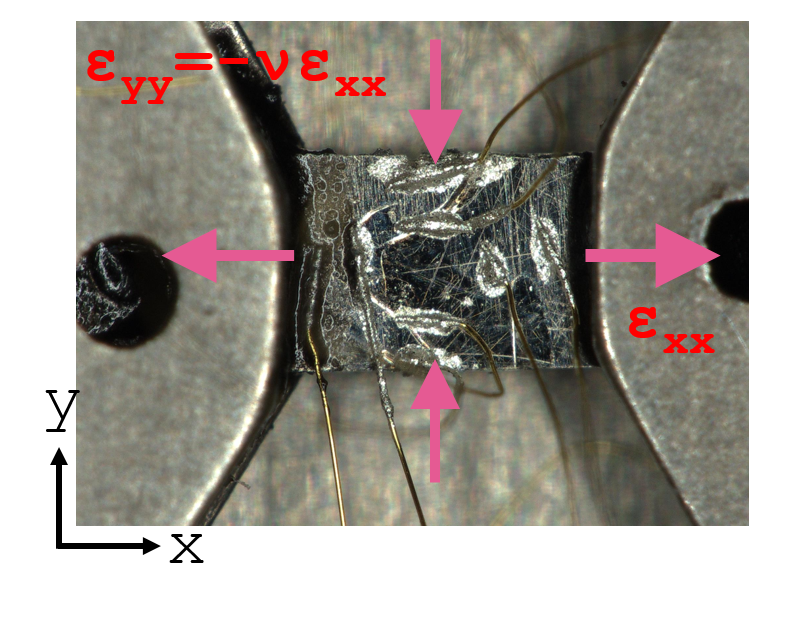}
	\caption{Photograph of a YbRh$_2$Si$_2$ crystal mounted on a CS-100 Razorbill strain device with four longitudinal and four transverse electrical contacts for elastoresistance measurements.}
	\label{figure1}
\end{figure}	

\section{Definitions}

Below we show how the longitudinal and transverse elastoresistance measurements can be combined to extract the elastoresitance coefficients along  the $A_{1g}$, $B_{1g}$, and $B_{2g}$ symmetry channels.
For uniaxial strain $\varepsilon$ applied along a given crystallographic direction, the measured fractional change in resistance is  
\begin{equation}
	\frac{\Delta R}{R_0}
	=
	\frac{\Delta \rho}{\rho_0}
	+
	(1 + 2\nu)\,\varepsilon
\end{equation}
The first term of Eq.~S1 describes the intrinsic contribution to elastoresistance and second term is due to the contribution from geometry. For strain applied along $x$ (e.g., $\varepsilon_{xx} = 100 $), the longitudinal and transverse strain components are
\begin{equation*}
	\varepsilon_{xx} = \varepsilon
	\qquad
	\varepsilon_{yy} = \varepsilon_{zz} = -\nu\,\varepsilon.
\end{equation*}
The resulting area change is
\begin{equation*}
	\frac{\Delta A}{A} \simeq \varepsilon_{yy} + \varepsilon_{zz}
	= -2\nu\,\varepsilon
\end{equation*}
leading to a geometric contribution
\begin{equation*}
	\left(\frac{\Delta R}{R_0}\right)_{\mathrm{geom}}
	= \frac{\Delta L}{L} - \frac{\Delta A}{A}
	= \varepsilon + 2\nu\varepsilon
	= (1 + 2\nu)\,\varepsilon
\end{equation*}

In tetragonal symmetry with applied strain along 100:

\[
\text{A}_{1g}\ \text{strain (isotropic):} \qquad
\varepsilon_{A_{1g}}
= \frac{1}{2}(\varepsilon_{xx} + \varepsilon_{yy})
= \frac{1}{2}(\varepsilon - \nu \varepsilon)
= \frac{1}{2}(1 - \nu)\varepsilon
\]

\[
\text{B}_{1g}\ \text{strain (x--y):} \qquad
\varepsilon_{B_{1g}}
= \frac{1}{2}(\varepsilon_{xx} - \varepsilon_{yy})
= \frac{1}{2}(\varepsilon + \nu \varepsilon)
= \frac{1}{2}(1 + \nu)\varepsilon
\]

\[
\text{B}_{2g}\ \text{strain:} \qquad
\varepsilon_{B_{2g}} = \varepsilon_{xy} = 0
\]
Uniaxial strain applied along $[100]$ decomposes into a mixture of $A_{1g}$ and $B_{1g}$ symmetry components. The corresponding elastoresistive response is
\begin{equation*}
	\frac{\Delta \rho}{\rho_0}
	= m_{A_{1g}}\varepsilon_{A_{1g}}
	+ m_{B_{1g}}\varepsilon_{B_{1g}} 
\end{equation*}
\begin{equation*}
	\frac{\Delta \rho}{\rho_0}
	= m_{A_{1g}}\!\left[\frac{1}{2}(1-\nu)\varepsilon\right]
	+ m_{B_{1g}}\!\left[\frac{1}{2}(1+\nu)\varepsilon\right]
\end{equation*}
\begin{equation*}
	\frac{\Delta \rho}{\rho_0}
	= \varepsilon
	\left[
	m_{A_{1g}}\frac{1-\nu}{2}
	+
	m_{B_{1g}}\frac{1+\nu}{2}
	\right]
\end{equation*}

Including both geometry factor and intrinsic response,  longitudinal change in resistance under $[100]$ strain,
\begin{equation*}
	\frac{d}{d\varepsilon_{[100]}}
	\!\left(\frac{\Delta R}{R_0}\right)_{[100]}
	=
	(1+2\nu_{[100]})
	+
	m_{A_{1g}}\!\left(\frac{1-\nu_{[100]}}{2}\right)
	+
	m_{B_{1g}}\!\left(\frac{1+\nu_{[100]}}{2}\right)
\end{equation*}
while the corresponding transverse response is
\begin{equation*}
	\frac{d}{d\varepsilon_{[100]}}
	\!\left(\frac{\Delta R}{R_0}\right)_{[010]}
	=
(1+2\nu_{[100]})
	+
	m_{A_{1g}}\!\left(\frac{1-\nu_{[100]}}{2}\right)
	-
	m_{B_{1g}}\!\left(\frac{1+\nu_{[100]}}{2}\right)
\end{equation*}
The geometric factor $(1+2\nu_{[100]})$ is therefore identical in both channels and cancels in the subtraction of the symmetry combinations as given by Eq.~S3. For the $A_{1g}$ (sum) combination, the geometric gauge factor does not cancel.
Purely geometric contribution $(1+2\nu)$ enters the longitudinal and
transverse responses with the same sign and therefore adds in the sum. Thus,
to obtain the intrinsic $A_{1g}$ elastoresistive coefficient one must either
(i) explicitly subtract the geometric term using $\nu$, or (ii) define
$m_{A_{1g}}$ from the outset to denote the geometry-subtracted (intrinsic)
quantity. After this subtraction, the remaining dependence on $\nu$ appears
only through the prefactor $1/(1-\nu)$ that converts the applied uniaxial
strain to the symmetry-resolved $A_{1g}$ strain component; this factor is not
a geometric contribution.
Combining the longitudinal and transverse responses derived above, we can
extract the symmetry-resolved elastoresistivity coefficients.  Hence, the elastoresistance coefficient along the symmetry chanel $m_{A_{1g}}$ and $m_{B_{1g}}$ are:
\begin{equation}
	m_{A_{1g}} = \frac{1}{1-\nu_{[100]}}
	\left[
	\frac{d(\Delta R/R_0)_{[100]}}{d\varepsilon_{[100]}}
	+
	\frac{d(\Delta R/R_0)_{[010]}}{d\varepsilon_{[100]}}
	\right]
		\label{eq:mA1g_100}
\end{equation}
\begin{equation}
	m_{B_{1g}} = \frac{1}{1+\nu_{[100]}}
	\left[
	\frac{d(\Delta R/R_0)_{[100]}}{d\varepsilon_{[100]}}
	-
	\frac{d(\Delta R/R_0)_{[010]}}{d\varepsilon_{[100]}}
	\right]
	\label{eq:mB1g}
\end{equation}

An entirely analogous procedure applies for uniaxial strain along $[110]$,
where the response mixes $A_{1g}$ and $B_{2g}$ symmetries. The longitudinal and transverse change in resistance for uniaxial strain along $[110]$ are:
\begin{align*}
	\frac{d}{d\varepsilon_{[110]}}
	\!\left(\frac{\Delta R}{R_0}\right)_{[110]}
	&=
	(1+2\nu_{[110]})
	+
	m_{A_{1g}}\!\left(\frac{1-\nu_{[110]}}{2}\right)
	+
	m_{B_{2g}}\!\left(\frac{1+\nu_{[110]}}{2}\right)
	\\
	\frac{d}{d\varepsilon_{[110]}}
	\!\left(\frac{\Delta R}{R_0}\right)_{[\bar{1}10]}
	&=
	(1+2\nu_{[110]})
	+
	m_{A_{1g}}\!\left(\frac{1-\nu_{[110]}}{2}\right)
	-
	m_{B_{2g}}\!\left(\frac{1+\nu_{[110]}}{2}\right)
\end{align*}
After subtracting the common geometric term, summing and subtracting results the elastoresistance coefficients for the symmetry channel $m_{A_{1g}}$ and $m_{B_{2g}}$ are:
\begin{equation}
	m_{A_{1g}} = \frac{1}{1-\nu_{[110]}}
	\left[
	\frac{d(\Delta R/R_0)_{[110]}}{d\varepsilon_{[110]}}
	+
	\frac{d(\Delta R/R_0)_{[\bar{1}10]}}{d\varepsilon_{[110]}}
	\right]
	\label{eq:mA1g_110}
\end{equation}
\begin{equation}
	m_{B_{2g}} = \frac{1}{1+\nu_{[110]}}
	\left[
	\frac{d(\Delta R/R_0)_{[110]}}{d\varepsilon_{[110]}}
	-
	\frac{d(\Delta R/R_0)_{[\bar{1}10]}}{d\varepsilon_{[110]}}
	\right].
	\label{eq:mB2g}
\end{equation}
Equations~\ref{eq:mA1g_100} and \ref{eq:mA1g_110} are the relations providing two independent determinations of $m_{A_{1g}}$,
serving as an internal consistency check on the analysis.
\section{In-plane poisson's ratio}
For a tetragonal crystal of symmetry , the elastic stiffness tensor in Voigt notation is

\begin{equation*}
	C =
	\begin{pmatrix}
		C_{11} & C_{12} & C_{13} & 0      & 0      & 0 \\
		C_{12} & C_{11} & C_{13} & 0      & 0      & 0 \\
		C_{13} & C_{13} & C_{33} & 0      & 0      & 0 \\
		0      & 0      & 0      & C_{44} & 0      & 0 \\
		0      & 0      & 0      & 0      & C_{44} & 0 \\
		0      & 0      & 0      & 0      & 0      & C_{66}
	\end{pmatrix}
	\label{eq:tetragonalC}
\end{equation*}

For a tetragonal  system with independent stiffness coefficients  Poisson's ratios along 100 and 110 are:
\[
\nu_{[100]} =
\frac{C_{13}^{2} - C_{12}C_{33}}
{C_{13}^{2} - C_{11}C_{33}}
\]

\[
\nu_{[110]} =
\frac{
	C_{33}\!\left(C_{11}+C_{12}-2C_{66}\right) - 2C_{13}^{2}
}{
	C_{33}\!\left(C_{11}+C_{12}+2C_{66}\right) - 2C_{13}^{2}
}
\]
The elastic tensor for YbRh$_2$Si$_2$ has been obtained from 
density-functional theory (DFT) within the generalized-gradient 
approximation \cite{deJong2015}. The obtained values (in GPa) form the 
following Voigt stiffness matrix:
\[
C_{ij} =
\begin{pmatrix}
	229 & 104 & 83  & 0   & 0   & 0 \\
	104 & 229 & 83  & 0   & 0   & 0 \\
	83  & 83  & 212 & 0   & 0   & 0 \\
	0   & 0   & 0   & 69  & 0   & 0 \\
	0   & 0   & 0   & 0   & 69  & 0 \\
	0   & 0   & 0   & 0   & 0   & 116
\end{pmatrix}
\]

The elastic tensor yields a Poisson's ratio of approximately $\nu_{[100]} \approx 0.36$  $[100]$, and $\nu_{[110]} \approx 0.07$.

\section{Effect of out-of-plane ($c$-axis) compression}

The expressions in Eqs.~(S2-S5) were obtained under the assumption that
only in-plane strain components contribute to the elastoresistance.  This
approximation is natural for an in-plane transport geometry and is sufficient
to establish the symmetry decomposition and the extraction procedure for the
$A_{1g}$, $B_{1g}$, and $B_{2g}$ channels.  However, a uniaxial in-plane strain
($\varepsilon_{\parallel}= \varepsilon_{[100]}$ or $\varepsilon_{[110]}$) in a tetragonal crystal
inevitably generates an accompanying out-of-plane strain $\varepsilon_{zz}$
through the directional Poisson ratio $\nu'$, defined as
$\nu'=-\varepsilon_{zz}/\varepsilon_{\parallel}$.
This out-of-plane deformation is fully symmetric and therefore transforms as
$A_{1g}$, contributing an additional isotropic strain component that must be
included in the complete elastoresistance response.

For a uniaxial in-plane strain $\varepsilon_{\parallel}$, the symmetric in-plane
$A_{1g}$ strain appearing in Eqs.~(S2) and (S4), given by
$(1-\nu)\varepsilon_{\parallel}/2$, must be combined with the out-of-plane
symmetric contribution $-\nu'\varepsilon_{\parallel}$. The total $A_{1g}$
strain entering the elastoresistance is thus the sum of these two
contributions.

\begin{equation*}
	\varepsilon_{A_{1g}}^{\mathrm{(total)}}=
	\left(
	\frac{1-\nu}{2}-\nu'
	\right)\varepsilon_{\parallel}
	\label{eq:A1g_total_modified}
\end{equation*}

For uniaxial strain along $[100]$ the longitudinal and transverse change in resistance including $c$ axis compression are now expressed as 
\begin{align*}
	\frac{d}{d\varepsilon_{[100]}}
	\left(\frac{\Delta R}{R_0}\right)_{[100]}
	&=
	(1+2\nu_{[100]})
	+
	m_{A_{1g}}
	\left(
	\frac{1-\nu_{[100]}}{2}-\nu'_{[100]}
	\right)
	+
	m_{B_{1g}}
	\left(
	\frac{1+\nu_{[100]}}{2}
	\right),
	\label{eq:L_modified}
	\\[4pt]
	\frac{d}{d\varepsilon_{[100]}}
	\left(\frac{\Delta R}{R_0}\right)_{[010]}
	&=
	(1+2\nu_{[100]})
	+
	m_{A_{1g}}
	\left(
	\frac{1-\nu_{[100]}}{2}-\nu'_{[100]}
	\right)
	-
	m_{B_{1g}}
	\left(
	\frac{1+\nu_{[100]}}{2}
	\right)
\end{align*}

Summing and subtracting these expressions now yields
\begin{equation*}
	\frac{d(\Delta R/R_0)_{[100]}}{d\varepsilon_{[100]}}
	+
	\frac{d(\Delta R/R_0)_{[010]}}{d\varepsilon_{[100]}}
	=
	2(1+2\nu_{[100]})
	+
	2 m_{A_{1g}}
	\left(
	\frac{1-\nu_{[100]}}{2}-\nu'_{[100]}
	\right)
\end{equation*}
\begin{equation*}
	\frac{d(\Delta R/R_0)_{[100]}}{d\varepsilon_{[100]}}
	-
	\frac{d(\Delta R/R_0)_{[010]}}{d\varepsilon_{[100]}}
	=
	2 m_{B_{1g}}
	\left(
	\frac{1+\nu_{[100]}}{2}
	\right)
\end{equation*}
Demonstrating explicitly that the $B_{1g}$ and $B_{2g}$ channel is unaffected by the
out-of-plane Poisson response, while excluding the geometric contribution the elastoresistance coefficient in $A_{1g}$ becomes

\begin{equation}
	m_{A_{1g}}
	=
	\frac{
		\dfrac{d(\Delta R/R_0)_{[100]}}{d\varepsilon_{[100]}}
		+
		\dfrac{d(\Delta R/R_0)_{[010]}}{d\varepsilon_{[100]}}
	}
	{1-\nu_{[100]} - 2\nu'_{[100]}}
	\label{eq:mA1g_modified}
\end{equation}
Comparison with Eq.~(S2) shows that the only change required to account for
$c$-axis compression is the replacement
\[
1-\nu_{[100]}
\;\longrightarrow\;
1-\nu_{[100]} - 2\nu'_{[100]}
\]

An entirely analogous modification applies for uniaxial strain along
$[110]$, where the in-plane response mixes $A_{1g}$ and $B_{2g}$ symmetries,
leaving the $B_{2g}$ channel unchanged but renormalizing the $A_{1g}$ term in
precisely the same manner as for applied strain along $[100]$.

Because both the in-plane symmetric strain and the Poisson-driven out-of-plane
strain transform as $A_{1g}$, their contributions enter with the same symmetry
and cannot be disentangled experimentally.  Thus our measured $m_{A_{1g}}$ response therefore
contains the combined response to symmetric in-plane biaxial strain and the
accompanying unavoidable $c$-axis compression.

\section{Effect of Reduced Strain-Transfer Efficiency}
The elastoresistance coefficients reported here were obtained using the
displacement measured by the capacitance sensor of the CS-100 strain cell.
Because the strain cell has finite mechanical compliance, part of the actuator
motion is accommodated by elastic deformation and a small rotation of the cell
rather than by uniform deformation of the sample itself. As a result, the
displacement inferred from the capacitance sensor slightly overestimates the
true strain experienced by the crystal. This effect can be captured by
introducing an effective strain-calibration factor $\eta<1$, such that the
actual strain in the sample is reduced relative to the nominal applied strain.
Consequently, the measured elastoresistance coefficients underestimate the true
values by a constant factor $1/\eta$. Since this correction arises from the
static mechanical properties and geometry of the strain cell, $\eta$ is
independent of temperature and magnetic field, so that all elastoresistance
curves are uniformly rescaled while their relative temperature and field
dependences remain unchanged.
The elastoresistance slopes shown in Fig.~3(a) were extracted under the
assumption of 100\% strain transfer between the piezoelectric actuator and the
YbRh$_2$Si$_2$ crystal.  If the actual strain transferred to the sample is
reduced, the extracted coefficients must be renormalized.  For a
strain-transfer efficiency $\eta$, the strain in the crystal is
$\varepsilon_{\mathrm{sample}}=\eta\,\varepsilon_{\mathrm{applied}}$, so that elastoresistance coefficient
\begin{equation*}
	m_{\mathrm{true}}(T,B)
	=
	\frac{d(\Delta R/R_0)}{d\varepsilon_{\mathrm{sample}}}
	=
	\frac{1}{\eta}
	\frac{d(\Delta R/R_0)}{d\varepsilon_{\mathrm{applied}}}
	=
	\frac{m_{\mathrm{meas}}(T,B)}{\eta}.
	\label{eq:stranfer_general}
\end{equation*}
$\eta $  is expected to be between 0.6–0.8 for the utilized strain device \cite{Hicks2014Rsi}; the exact value does not affect conclusions. For instance $\eta = 0.70$ results
\begin{equation*}
	m_{\mathrm{true}}(T,B) \simeq 1.43\,m_{\mathrm{meas}}(T,B).
	\label{eq:transfer_70percent}
\end{equation*}

Thus, every curve in Fig.~3(a) would be uniformly rescaled upward by a factor
$1/\eta$, while the temperature and field dependence of the data remain
unchanged. Only the absolute magnitude of the elastoresistivity is modified
by the reduced strain transfer.  Because the rescaling is uniform, the
relative enhancement of $m(T)$ is unaffected. For example, the increase from
$m \sim 2$ at high temperature to $m \sim 45$ at low temperature simply
becomes $m \sim 3$ to $m \sim 64$ for $\eta=0.70$, leaving the percent change
unchanged.  Uncertainties in strain-transfer efficiency therefore alter only
the overall scale of $m(T)$ and do not influence the qualitative or
quantitative trends discussed in the text.

\section{Linearity of the Elastoresistance at Large Strain}

\begin{figure}[h]
	\includegraphics[scale=0.35]{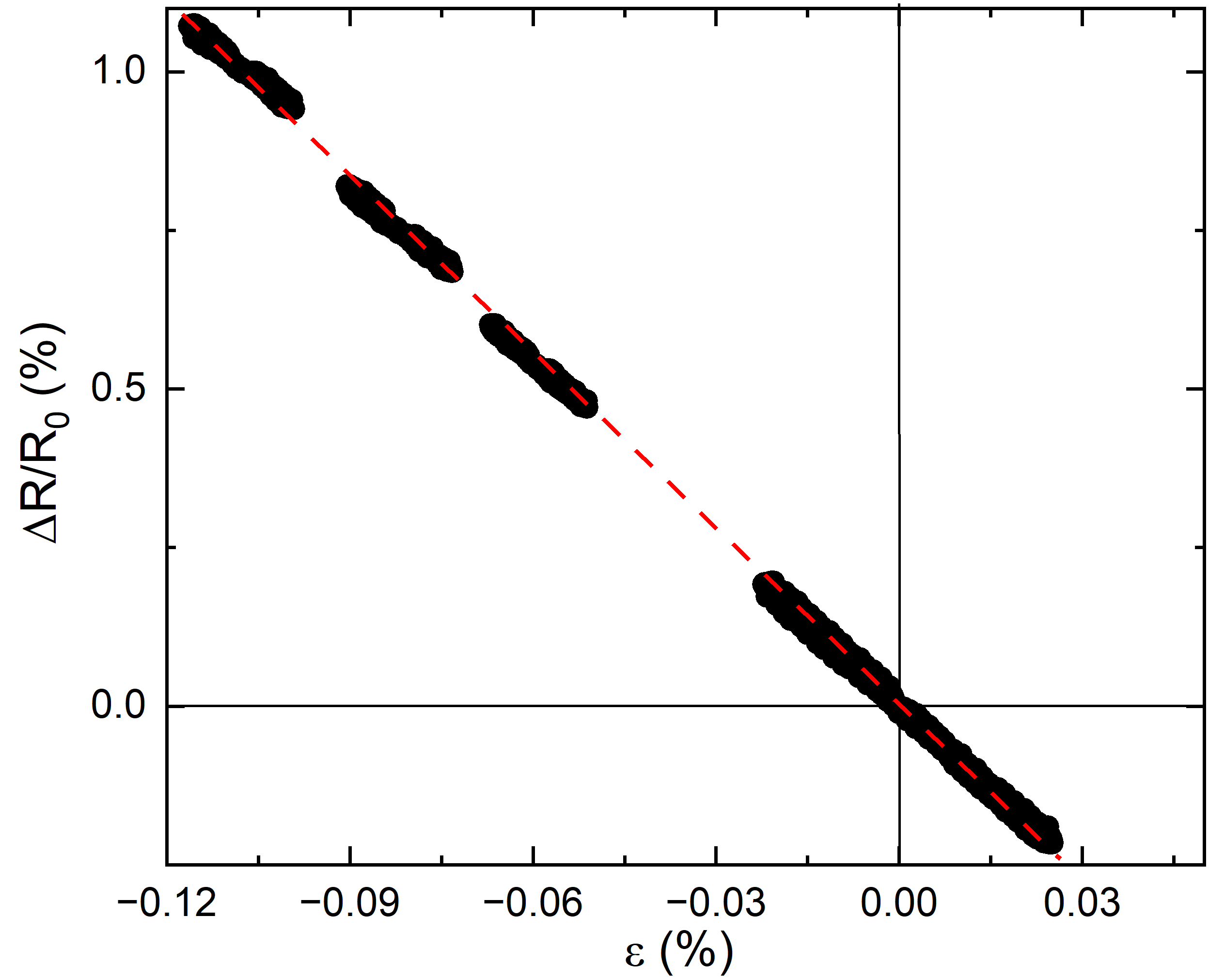}
	\caption{Percentage change in strain driven resistance $\Delta R/R_0$ as a function of uniaxial strain $\varepsilon$ measured along the same crystallographic direction as in the main text, shown over an extended strain range. The red dashed line is a linear fit to the data, demonstrating that the elastoresistance remains linear up to the maximum applied compressive strain of $\sim 0.12\%$. This extended-range plot confirms that the elastoresistance coefficients reported in the main text are not affected by nonlinear strain response.
	}
	\label{SMfigure4}
\end{figure}	

As discussed in the main text, elastoresistance measurements were generally
performed within a narrow strain window centered near the nominal zero-strain/strain neutral
point, typically $\varepsilon_{xx} = \pm 0.04\%$ [Fig.~2 of the main text]. Within this restricted strain range, the resistance varies strictly linearly
with strain, allowing the elastoresistance coefficients to be extracted
reliably from the slope. Here, the strain-neutral point is defined as the reference strain corresponding
to zero voltage applied to the piezostacks of the Razorbill device. Strain
sweeps are performed about this point, within a regime where the
resistance--strain response is locally linear and symmetric. This
strain-neutral point does not necessarily coincide with the \textit{true}
zero-strain condition, as it can be shifted by differential thermal
contraction during thermal cycling.
To verify that the extracted elastoresistance coefficients are not sensitive
to the choice of this reference point, we performed additional measurements at
$T = 80$~K over a much wider strain range. Specifically, small-amplitude strain
sweeps were carried out about multiple centre positions spanning a total range
exceeding compressive strain $\varepsilon_{xx} = -0.1\%$, as shown in Fig.~S2. At each center
position, the resistance remains linear over the local sweep range, and the
extracted slope is found to be independent of the absolute strain offset. This
demonstrates that the elastoresistance is governed by the local linear response
and does not depend on the precise value chosen as the zero-strain reference.

These measurements confirm that the elastoresistance coefficients reported in
the main text reflect an intrinsic linear response of the system and are not
affected by residual strain offsets or by the absolute strain level within the
investigated range. The restriction to small strain amplitudes in the main text
therefore ensures operation well within the linear-response regime while
yielding results that are representative of the broader strain behavior.

\section{Magnetic field dependence of the longitudinal elastoresistance}

	\begin{figure}[h]
	\includegraphics[scale=0.40]{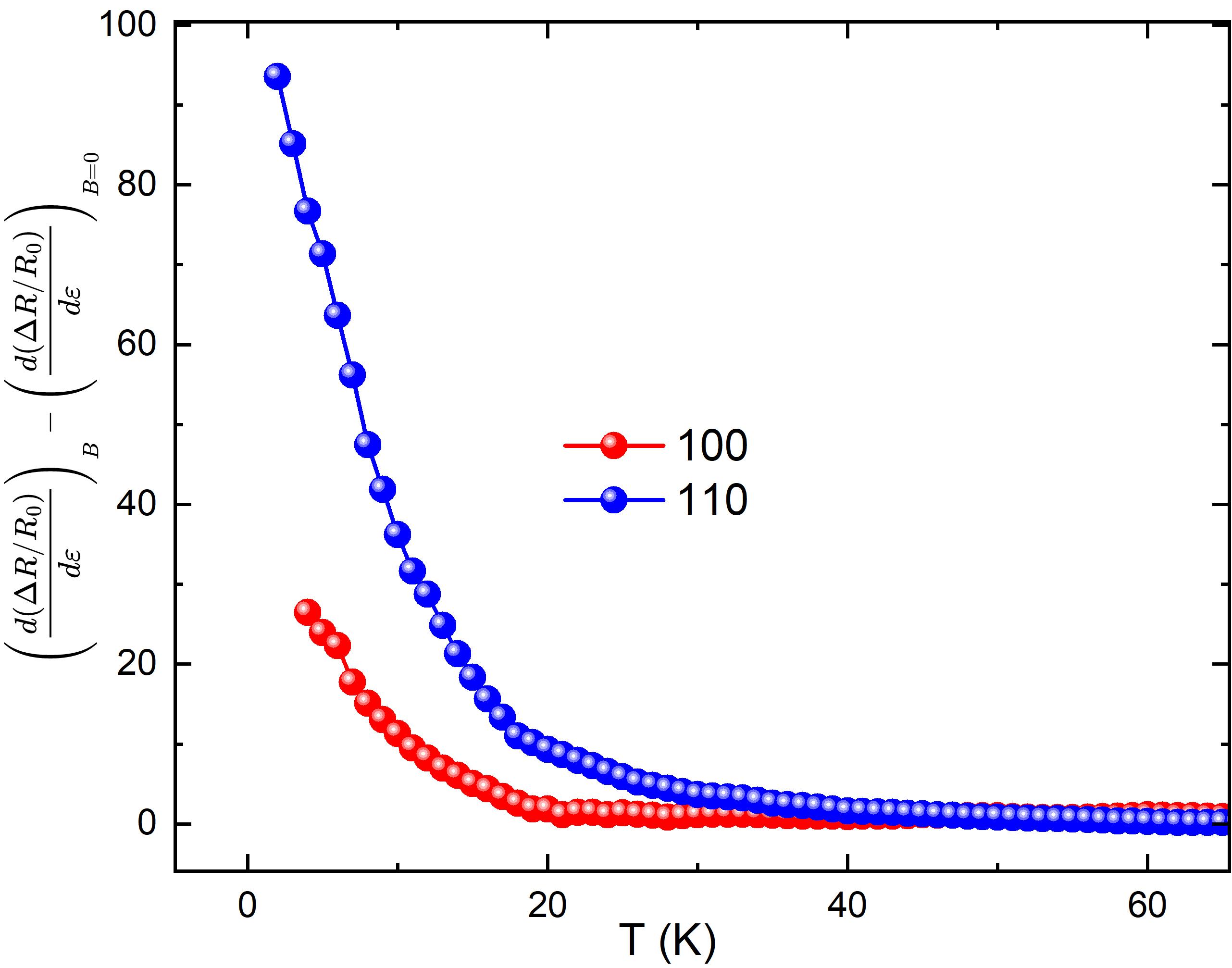}
	\caption{Temperature dependence of the relative longitudinal resistance change under strain $d(\Delta R/R_0)/d\varepsilon$ in finite magnetic field of 14~T and 13~T for strain along the [100] and [110] directions, respectively, minus the respective measurements in zero field. Resistance was measured parallel to the strain and applied magnetic field.}
	\label{SMfigure3a}
\end{figure}	

Fig.~\ref{SMfigure3a} displays the difference between the longitudinal elastoresistance in magnetic fields of 14 T (for $\epsilon \parallel [100]$) and 13 T ($\epsilon \parallel [110]$) and respective zero-field data (shown in the main text Fig.~3(a)) as function of temperature. Please note, that we have not done transverse elastoresistance measurements in magnetic field.  Therefore a full symmetry analysis of the elastoresistance in magnetic field has not been possible. For the further discussion on the impact of longitudinal magnetic fields on the elastoresistance, we refer to the main text.

	\begin{figure}[h]
	\includegraphics[scale=1.0]{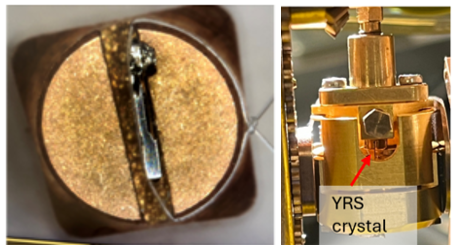}
	\caption{(a) YbRh$_2$Si$_2$ single crystal secured with a PTFE thread and mounted on a slotted sample stage for measurements using a capacitive dilatometer. Such mounting  facilitates the handling and reliable positioning of thin, fragile crystals while ensuring good mechanical stability during measurement.
		(b) The same YbRh$_2$Si$_2$ crystal mounted inside the capacitive dilatometer prior to measurement, illustrating the final experimental configuration.}
	\label{SMfigure2}
\end{figure}	

	\begin{figure}[h]
	\includegraphics[scale=0.40]{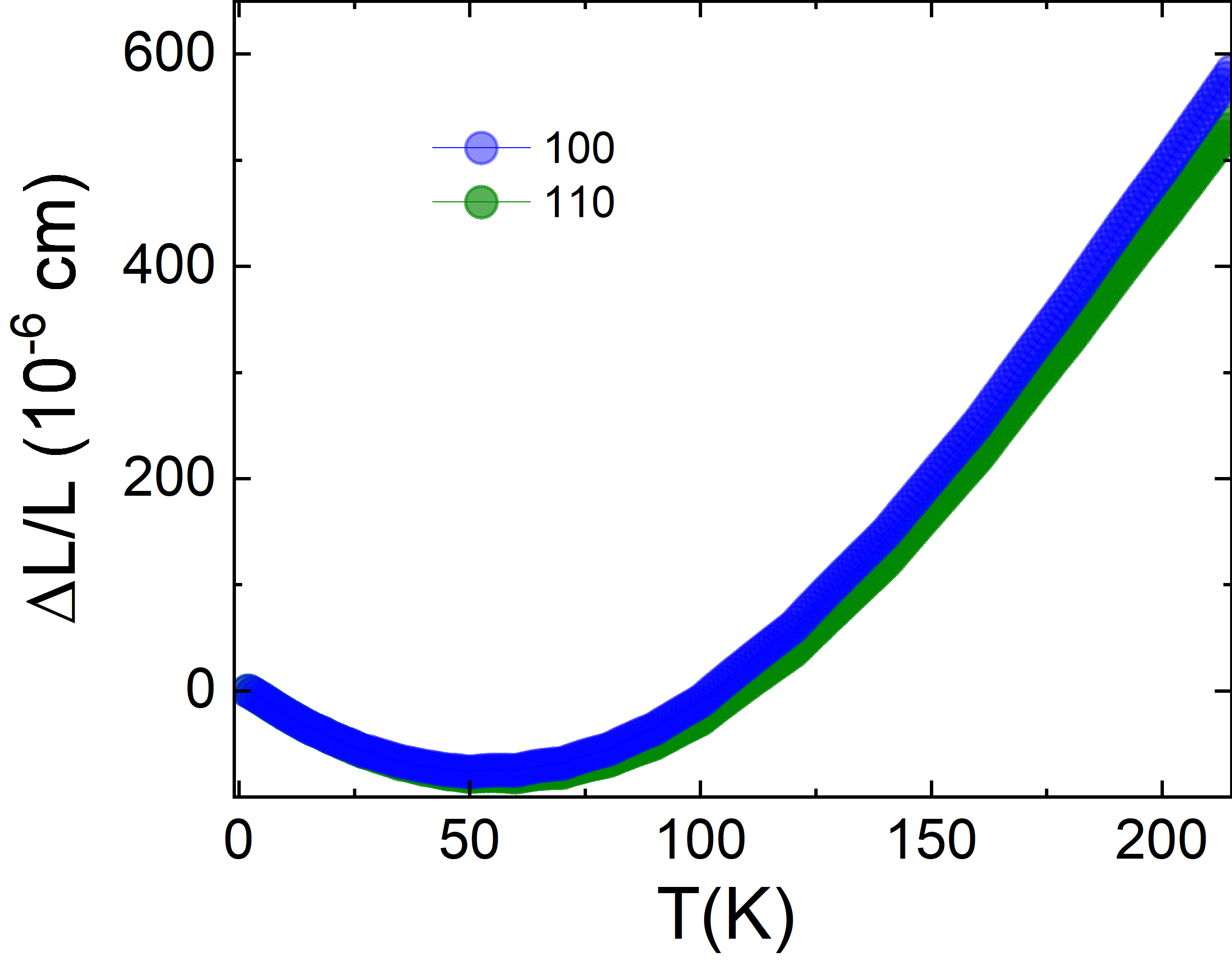}
	\caption{The relative length change $\Delta L/L_0$ of a YbRh$_2$Si$_2$ single crystal along 100 and 110 crystallographic direction.
	}
	\label{SMfigure3}
\end{figure}

\section{In-plane thermal expansion measurement on thin YbRh$_2$Si$_2$ platelets}

Capacitive dilatometry provides high-resolution access to temperature induced length changes along selected crystallographic directions. In capacitive dilatometers, a  specimen is positioned against the movable electrode and secured using a mechanical clamping mechanism \cite{Kuchler2012}. This configuration requires samples that are sufficiently rigid and self-supporting, typically with thickness of millimeter-scale dimensions, in order to stand the crystal upright and mechanical stability during installation. As a result, extremely thin or mechanically delicate samples cannot be mounted reproducibly using standard methods  with thicknesses of the order of a few hundred of micrometers. Here, we employ our recently developed novel sample-mounting strategy \cite{Panja2025} using slotted mounting stamp (see Fig.~S3) that overcomes these constraints and, for the first time, enables direct reliable measurements of the thermal expansion along the in-plane $100$ and $110$ crystallographic directions of a thin platelet single crystal of YbRh$_2$Si$_2$. \\
In a capacitive dilatometer, the measured length change contains not only the intrinsic thermal expansion of the sample but also a contribution from the  dilatometer, arising from the thermal expansion of the Be-Cu cell body itself. To isolate the sample response, a background subtraction procedure is required. The cell contribution is determined by combining a direct measurement of the empty-cell response with literature thermal expansion data of copper \cite{Kuchler2012}. The intrinsic relative length change of the sample is then obtained as
\begin{equation}
	\left(\frac{\Delta L}{L_0}\right)_{\mathrm{sample}}
	=
	\frac{\Delta L^{\mathrm{meas}}_{\mathrm{sample}}(T)
		-
		\Delta L_{\mathrm{empty\ cell}}(T)}{L_0}
	+
	\left(\frac{\Delta L}{L}\right)^{\mathrm{Cu}}_{\mathrm{lit}}(T),
	\label{eq:background_subtraction}
\end{equation}
where $\Delta L^{\mathrm{meas}}_{\mathrm{sample}}$ is the measured response of the sample, $\Delta L_{\mathrm{empty\ cell}}$ accounts for the  background of empty cell, and $(\Delta L/L)^{\mathrm{Cu}}_{\mathrm{lit}}$ is the literature thermal expansion of copper. The empty-cell contribution is determined from measurements performed on a pure copper block with the same length as the YbRh$_2$Si$_2$ crystal. The in-plane relative length along the $100$ and $110$ crystallographic directions are shown in Fig.~S4.

The linear thermal expansion coefficient $\alpha_i=\frac{d (\Delta L_i/L_{i,0})}{dT}$ along the direction $i$ is obtained by calculating the numerical derivative of the normalized length change with temperature. Thermodynamically, the linear thermal expansion

\begin{equation*}
 \alpha_i=-(V_m)^{-1}\frac{\partial S}{\partial p_i} =(V_mY)^{-1} \frac{\partial S}{\partial \epsilon_i}
\end{equation*} 

  probes the symmetry-conserving uniaxial pressure derivative of entropy $S$ along the direction $i$, where $V_m$ is the molar volume, which can also be expressed as symmetric strain derivative of the entropy, using  Young's modulus $Y$.

Following~\cite{Wiecki2021}, we derive a relation between the expansion coefficient and the symmetric elastoresistance, based upon the Fisher-Langer scaling relation for critical magnetic scattering in metallic magnets

\begin{equation*}
	\frac{\partial \rho}{\partial T} \propto C_{\mathrm{mag}},
\end{equation*}

which relates the temperature derivative of the electrical resistance to the magnetic specific heat ~\cite{Fisher1968}.

The temperature derivative of the linear thermal expansion 

\begin{equation*}
	\frac{d\alpha_i}{dT}=(V_mY)^{-1} \frac{\partial^2 S}{\partial \epsilon_i \partial T}=(V_mY)^{-1}\frac{\partial}{\partial \epsilon_i}\frac{C}{T} \sim \frac{\partial}{\partial \epsilon_i}\frac{1}{T}\frac{\partial \rho}{\partial T}\sim\frac{1}{T}\frac{\partial (m_iR(T)/R_{\rm 300K})}{\partial T}
	\end{equation*}
	
can thereby be expressed by the temperature derivative of the symmetric elastoresistance. It follows that 

\begin{equation}
	\alpha_i \sim\int \frac{1}{T}\frac{\partial (m_iR(T)/R_{\rm 300K})}{\partial T} dT + const.
	\label{eq:fisher_ta}
	\end{equation}
	
the temperature dependence of thermal expansion follows that of the above integral.

\begin{figure}[h]
	\includegraphics[scale=0.40]{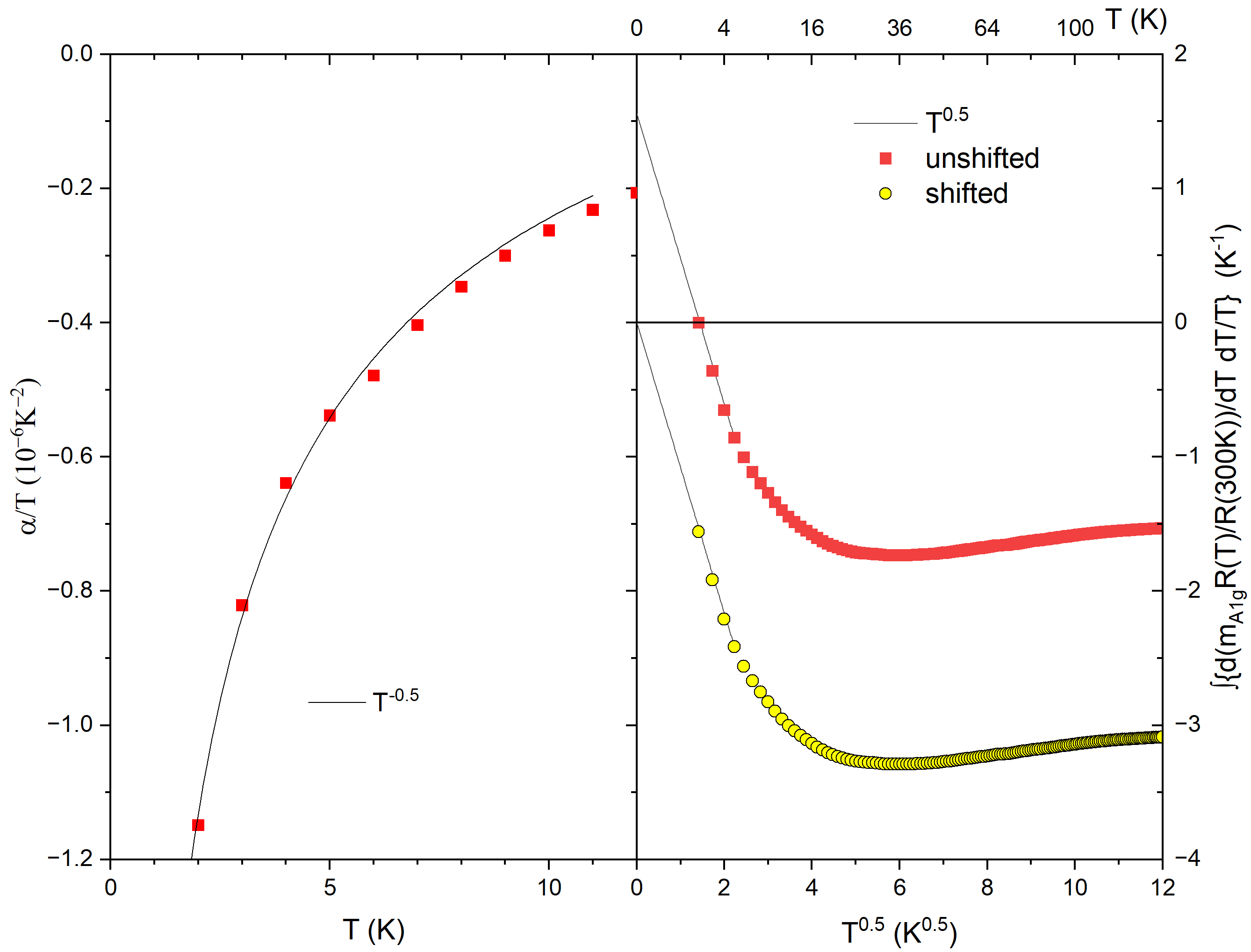}
	\caption{Left: linear thermal expansion of YbRh$_2$Si$_2$ along [100] as $\alpha/T$ vs. $T$. The solid line is a fit according to a $T^{-0.5}$ dependence. Right:	temperature dependence of the intetragal from Eq.(S8) (red squares), starting at a temperature of 2~K vs. $T^2$. The black line indicates a linear fit, representing a $T^2$ dependence of the integral, which would correspond to the $T^{-0.5}$ dependence of $\alpha/T$. The fit extrapolates to a value of 1.557 at $T=0$. The yellow circles display shifted data (the red quares are subtracted by 1.557), yielding a linear extrapolation to zero at $T=0$. Note, that for the [100] direction (not shown) a shift of 1.4 was used.}
	\label{SMfigure}
\end{figure}

Since our data start at a temperature of 2~K, the integral has a value of zero at 2~K, but there is no reason why at this temperature the entropy shall not have a strain dependence anymore. In fact, it is known that $\alpha/T$ diverges upon cooling in this temperature range and changes sign at $T_N=70$ mK~\cite{Kuechler2004,Kuechler2003}. Our $\alpha/T$ data measured in the PPMS down to 2~K display a $T^{-0.5}$ divergence below 10 K, as shown in the left panel of Figure~\ref{SMfigure}. Respectively, we fitted a square-root behavior to the ``Fisher thermal expansion'' of Eq.(S8) (which corresponds to $\alpha$). In the right panel of Figure~\ref{SMfigure} this property is shown on square-root temperature scale. The fit yields the black line through the red circles (from the measurement along [100]). The extrapolated value of 1.557 (1.4 for [110], not shown) at $T=0$ has then been subtracted (see yellow circles for [100]) for the plot of the ``Fisher thermal expansion'' divided by temperature, shown in Fig. 5 of the main text.


\end{document}